\documentclass[useAMS,usenatbib]{mn2e}

\usepackage{graphics}
\usepackage{graphicx}
\usepackage{psfig}
\input{epsf}

\title[Observational constraints on the braneworld model...]{Observational constraints on the braneworld model
\\ with brane-bulk energy exchange}
  \author[{M. Sadegh Movahed and Ahmad Sheykhi}]{M. Sadegh Movahed$^{1,2}$\thanks{E-mail:m.s.movahed@ipm.ir},
  Ahmad Sheykhi$^{3}$\thanks{E-mail:sheykhi@gmail.com}.\\ $^{1}$ Department of Physics, Shahid Beheshti University, Evin, Tehran 19839, Iran.\\
   $^{2}$School of Astronomy, IPM (Institute for Studies in theoretical Physics and Mathematics), P.O.Box 19395-5531,Tehran,
Iran.\\ $^{3}$Department of Physics, Shahid Bahonar University, P.O.
Box 76175-132, Kerman, Iran.}
\begin{document}

\date{Accepted . Received ; in original form }

\pagerange{\pageref{firstpage}--\pageref{lastpage}} \pubyear{2007}

\maketitle

\label{firstpage}

\begin{abstract}

We investigate the viability of the braneworld model with energy
exchange between the brane and bulk, by using the most recent
observational data related to the background evolution. We show that
this energy exchange behaves like a source of dark energy and can
alter the profile of the cosmic expansion. The new Supernova Type Ia
(SNIa) Gold sample, Supernova Legacy Survey (SNLS) data, the
position of the acoustic peak at the last scattering surface from
the Wilkinson Microwave Anisotropy Probe (WMAP) observations and the
baryon acoustic oscillation peak found in the Sloan Digital Sky
Survey (SDSS) are used to constrain the free parameters of this
model. To infer its consistency with the age of the Universe, we
compare the age of old cosmological objects with what computed using
the best fit values for the model parameters. At $68\%$ level of
confidence, the combination of Gold sample SNIa, Cosmic Microwave
Background (CMB) shift parameter and SDSS databases provide
$\Omega_m=0.29_{-0.02}^{+0.03}$, $\Omega_{A}=-0.71_{-0.03}^{+0.03}$
and $\mu=-0.40_{-0.26}^{+0.28}$, hence a spatially flat Universe
with $\Omega_K=0.00_{-0.04}^{+0.04}$. The same combination with SNLS
supernova observation give $\Omega_m=0.27_{-0.02}^{+0.02}$,
$\Omega_{A}=-0.74_{-0.02}^{+0.04}$ and $\mu=0.00_{-0.30}^{+0.30}$
consequently provides a spatially flat Universe
$\Omega_K=-0.01_{-0.03}^{+0.04}$. These results obviously seem to be
compatible with the most recent WMAP results indicating a flat
Universe.

\end{abstract}


\section{Introduction}
Recent observations of type Ia supernovas (SNIa) suggest the
expansion of the Universe is accelerating
\citep{ris,permul,R04,Tonry}. As it is well known all usual types of
matter with positive pressure generate attractive forces, which
decelerate the expansion of the Universe. A ``dark energy" component
with negative pressure was suggested to account for the invisible
fuel that drives the current acceleration of the Universe. Although
the nature of such dark energy is still speculative, an overwhelming
flood of papers has appeared which attempt to describe it by
devising a great variety of models (see
\citep{sa00,wein,lim04,cop06,arm00,pad03} for recent reviews).
Available models of dark energy differ in the value and variation of
the equation of state parameter, $w$, during the evolution of the
Universe. Among them are cosmological constant $\Lambda$, an
evolving scalar field (referred to by some as quintessence), the
phantom energy, in which the sum of the pressure and energy density
is negative , the quintom model, the holographic dark energy , the
Chaplygin gas, and the Cardassion model. Another approach dealing
with this problem is using the modified gravity by changing the
Einstein-Hilbert action. Some of models as $1/R$ and logarithmic
models provide an acceleration for the Universe at the present time
\citep{wein,carr,pee,padm,bennett,peri,spe03,sa1,sa2,wet88,Ratra,Friem,Turn,Caldw,Liddle,Zlate,stei,Torres,peb88,cal03,arb05,wan00,per99,pag03,dor01,dor04,cal03,cal02,da03,LAmendola2000,LAmendola2001,amen3,piet3,come3,fran4,zhang5,zong,li04,Wang1,Wang2,Wang3,be02,kam01,zong,clif,da04,noji1,noji2,Deff,freese,ahmad,arkani,Dvali02,rahvar07,sar07}.

Independent of these challenges, we deal with the dark energy
puzzle. In recent years, theories of large extra dimensions, in
which the observed Universe is realized as a brane embedded in a
higher dimensional spacetime, have received a lot of interest.
According to the braneworld scenario, the standard model of particle
fields are confined to the brane while, in contrast, the gravity is
free to propagate in the whole spacetime \citep{RSII,Dvali}. In
these theories the cosmological evolution on the brane is described
by an effective Friedmann equation that incorporates non-trivially
with the effects of the bulk into the brane \citep{Bin,She1,She20}.
An interesting consequence of the braneworld scenario is that it
allows the presence of five-dimensional matter which can propagate
in the bulk space and may interact with the matter content in the
braneworld. It has been shown that such interaction can alter the
profile of the cosmic expansion and leads to a behavior that would
resemble the dark energy. The cosmic evolution of the braneworld
models with energy exchange between the brane and bulk has been
studied in the different approaches
 \citep{Kirit1,Kirit2,apos1,apos2,
Kirit3,Umezu,Kof,Cai,Bog1,Bog2,Sheykhi,ghkhm}.


In the framework of the braneworld scenarios, many attempts to
observationally detect or distinguish brane effects, on the
evolution of our Universe, from the usual dark energy physics have
been discussed in the literature \citep{Pires,Capo04}. In
\citep{Sah1,Sah2} a class of braneworld models has been
investigated. A new and interesting feature of this class of models
is that the acceleration of the Universe may be a transient
phenomenon, which cannot be achieved in the context of our current
standard scenario, i.e., the $\Lambda$CDM model but could reconcile
the supernova evidence for an accelerating Universe with the
requirements of string/M-theory \citep{Fis}. The purpose of the
present work is to disclose the effect of energy exchange between
the brane and bulk in Randall-Sundrum II braneworld scenario on the
evolution of the Universe. Giving the wide range of cosmological
data available, we are able to test the viability of this class of
braneworld models by putting recent observational constraints on its
free parameters.

We have three independent types of observational constraints for the
dark energy models: (\textit{i}) the supernova distance modulus
\citep{ris,R04,Per1,Nob05}, (\textit{ii}) the dynamical evidence for
matter density \citep{spe03} and (\textit{iii}) the age of the
Universe \citep{Knox,Hu}. Besides, a great success has been scored
in high precision measurements of CMB anisotropy, as well as in
galaxy clustering \citep{Wein2,Selj,refre,Heyman}. Among these
observations, the age of the Universe is one of the most pressing
pieces of data disclosing information about dark energy. Indeed, any
limit on the age of the Universe during its evolution with redshift
will reveal the nature of dark energy. This is due to the fact that
dark energy influences the evolution of the Universe. However,
different models of dark energy may lead to the same age of Universe
at $z = 0$. To lift this degeneracy, we should examine the age of
the Universe at different stages of its evolution and compare it
with the estimated age of high-redshift objects. This procedure
constrains the age at different stages, being a powerful tool to
test the viability of different models \citep{Wang2,Fri}.

This paper is organized as follows: In section $2$, we introduce a
braneworld model with energy exchange between the brane and bulk,
the cosmology of this model, its free parameters and background
dynamics of the Universe governed by the effective Friedmann
equation. We also show how this model can exhibit acceleration
expansion of our Universe. Most limitations regarded to this
interaction in our model are introduced. We investigate the
geometrical effects of underlying braneworld cosmology in section
$3$. In section $4$, we test the viability of our model by putting
some constraints on the parameters of the model. For this aim, we
use the new Gold sample and Legacy Survey of SNIa data (Riess et al.
2004; Astier et al. 2005), its combination with the position of the
observed acoustic angular scale on CMB and the baryonic oscillation
length scale. In section $5$, we compare the age of the Universe in
this model with the age of old cosmological objects. The last
section is devoted to conclusions and discussions.

\section{Braneworld With Brane-Bulk Energy Exchange}

We start from the following action
\begin{eqnarray}\label{Act}
S &=& \frac{1}{2{\kappa}^2} \int{
d^5x\sqrt{-{g}}\left({R}-2\Lambda\right)} +\int{
d^5x\sqrt{-{g}}{L}_{bulk}^{m}}   \nonumber \\
&&+\int {d^{4}x\sqrt{-\tilde{g}}({L}_{brane}^{m}-\sigma)},
\end{eqnarray}
where $R$ is the 5D scalar curvature and $\Lambda<0$ is the bulk
cosmological constant. $g$ and $\tilde{g}$ are the bulk and the
brane metrics, respectively. Throughout this paper we choose the
unit $\kappa^2=1$ as the gravitational constant in five dimension.
We have also included arbitrary matter content both in the bulk and
on the brane through ${L}_{bulk}^{m}$ and $ {L}_{brane}^{m}$
respectively. $\sigma$ is the positive brane tension. The field
equations can be obtained by varying the action, equation
(\ref{Act}), with respect to the bulk metric $g_{AB}$. The result is
\begin{eqnarray}
G_{AB}+\Lambda g_{AB}= T_{AB} \label{Feq}.
\end{eqnarray}
For convenience we choose the extra-dimensional coordinate $y$ such
that the brane is located at $y=0$ and bulk has ${Z}_2$ symmetry. We
are interested in the cosmological solution with a metric
\begin{eqnarray}
ds^2&=&-n^2(t,y) dt^2 + a^2(t,y)\gamma_{ij}dx^i dx^j + b^2(t,y)dy^2,
\label{line}
\end{eqnarray}
where $\gamma _{ij}$ is a maximally symmetric $3$-dimensional metric
for the surface ($t$=const., $y$=const.), whose spatial curvature is
parameterized by $K = -1, 0, 1$. The metric coefficients $n$ and $b$
are chosen $n(t,0)=1$ and $b(t,0)=1$, where $t$ is cosmic time on
the brane. The total energy-momentum tensor has bulk and brane
components and can be written as
\begin{equation}
{T}_{AB}=
{T}_{AB}\mid_{brane}+{T}_{AB}\mid_{\sigma}+{T}_{AB}\mid_{bulk}.
\end{equation}
The first and the second terms are the contribution from the
energy-momentum tensor of the matter field confined to the brane and
the brane tension
\begin{eqnarray}
T^{A}_{\,\,B}\mid_{brane}\,&=&\,\mathrm{diag}(-\rho,p,p,p,0)\frac{\delta(y)}{b}{\label{bem}},\\
T^{A}_{\,\,B}\mid_{\sigma}\,&=&\,\mathrm{diag}(-\sigma,-\sigma,-\sigma,-\sigma,0)\frac{\delta(y)}{b}{\label{sigma}},
\end{eqnarray}
where $\rho$, and $p$, being the energy density and pressure on the
brane, respectively. In addition, we assume an energy-momentum
tensor for the bulk content with the following form
\begin{equation}\label{bulk}
T^{A}_{\ B}\mid_{bulk}\,= \,\left(\begin{array}{ccc}
T^{0}_{\ 0}\,&\,0\,&\,T^{0}_{\ 5}\\
\,0\,&\,T^{i}_{\ j}\delta^i_{\ j}\,&\,0\\
-\frac{n^2}{b^2}T^{0}_{\ 5}\,&\,0\,&\,T^{5}_{\ 5}
\end{array}\right)\,\,\,\,\,.
\end{equation}
The quantities which are of interest here are $T^{5}_{\ 5}$ and
$T^{0}_{\ 5}$, as these two enter the cosmological equations of
motion. In fact, $T^{0}_{\ 5}$ is the term responsible for energy
exchange between the brane and the bulk.  Integrating the $(00)$ and
the $(ij)$ components of the field equations (\ref{Feq}) across the
brane and imposing ${Z}_2$ symmetry, we have the jump across the
brane
\begin{eqnarray}\label{jun1}
\frac{a'_{+}}{ a_{0} }
 &=&-\frac{1}{6}(\rho+\sigma),\\
 \frac{n'_{+}}{n_{0}} &=&\frac{1}{6}(2\rho+3p-\sigma),\label{jun2}
\end{eqnarray}
where $ 2a'_{+}=-2a'_{-}$  and $ 2n'_{+}=-2n'_{-}$ are the
discontinuities of the first derivative and primes denote
derivatives with respect to $y$. In addition, as usual, the
subscript `` 0" denotes quantities  are evaluated at $y=0$.

Substituting the junction conditions i.e. equations $(\ref{jun1})$
and $(\ref{jun2})$ into the $(55)$ and $(05)$ components of the
field equation $(\ref{Feq})$, we obtain the modified Friedmann
equation and the semi-conservation law on the brane
\begin{eqnarray}\label{fri1}
2H^2+\dot{H}+\frac{K}{a_{0}^2}&=&-\frac{1}{36}\left[\sigma\left(3p-\rho\right)+\rho\left(\rho+3p\right)\right]
\nonumber \\
&&+\frac{1}{3}\left(\Lambda+\frac{\sigma^2}{6}\right)-\frac{T^{5}_{\
5}}{3},
\\
\dot{\rho}+3H(\rho+p)&=&-2 T^{0}_{\ 5},\label{T1}
\end{eqnarray}
where $H={\dot{a}_{0}}/{a_{0}}$ is the Hubble parameter on the brane
and dots denote time derivative. We shall assume an equation of
state $p=w\rho$ which represents a relation between the energy
density and pressure of the matter on the brane. The bulk matter
contributes to the energy content of the brane through the bulk
pressure terms $T^{5}_{ \ 5}$ and $T^{0}_{ \ 5}$. In order to derive
a solution that is largely independent of the bulk dynamics, we
should neglect $T^{5}_{ \ 5}$ term by assuming that the bulk matter
relative to the bulk vacuum energy is much less than the ratio of
the brane matter to the brane vacuum energy \citep{Kirit1}.
Considering this we get

\begin{eqnarray}\label{fri2}
&&2H^2+\dot{H}+\frac{K}{a_{0}^2}=\gamma
\rho\left(1-3w\right)-\beta\rho^2\left(1+3w\right)
+\frac{\lambda}{3},
\\
&&\dot{\rho}+3 H \rho(1+w)=-2 T^{0}_{\ 5},\label{T2}
\end{eqnarray}
where we have used the usual definition $\beta\equiv1/{36}$,
$\gamma\equiv \beta \sigma$ and
$\lambda\equiv(\Lambda+{\sigma^2}/{6})$. Assuming the
Randall-Sundrum fine-tuning $\lambda=\Lambda+\sigma^2/6=0$  holds on
the brane, one can easily check that the Friedmann equation
(\ref{fri2}) is equivalent to the following equations
 \begin{eqnarray}\label{fri3}
{H}^{2}+\frac{K}{a_{0}^2}=\beta\rho^2+2\gamma\rho+\chi,\\
\dot {\chi}+ 4\,H \chi=4T^0_{\ 5}(\beta\rho+\gamma).\label{chi}
 \end{eqnarray}
Equation (\ref{fri3}) is the modified Friedmann equation describing
cosmological evolution on the brane. The auxiliary field $\chi$
incorporates non-trivial contributions of dark energy which differ
from the standard matter fields confined to the brane. It is worth
noting that the flow of the mass-energy from the bulk onto the brane
may resemble as the dark energy. Indeed it can influence the
background evolution of the Universe and leads to acceleration (see
e.g. \citep{Kirit1}). One may argue that whether the energy exchange
between the brane and bulk becomes dark matter or not? To answer to
this question, one should consider an interaction between dark
matter and dark energy on the brane which is not clear yet. Besides
in order to have the equation of state in the bulk, a particular
model of the bulk matter is required which is not clear yet, because
we do not exactly know the bulk geometry \citep{Bog1}. So now in our
coarse-grained model we ignored this effect.

We are also interested in the scenarios where the energy density
of the brane is much lower than the brane tension, namely
$\rho\ll\sigma$, therefore equations (\ref{fri3}) and (\ref{chi})
can be simplified in the following form
\begin{eqnarray}\label{fri4}
{H}^{2}+\frac{K}{a_{0}^2}&=& 2\gamma \rho+\chi,\\
 \dot {\chi}+ 4\,H \chi&=&4 \gamma \ {T^0_{\ 5}}.\label{chi4}
\end{eqnarray}
Then we take ansatz ${T}^{0}_{\ 5}= A H a^{\mu}$ for the brane-bulk
energy exchange \citep {Cai}, where $A$ and $\mu$ are arbitrary
constants and thereafter we have omitted the ``0" subscript from the
scale factor on the brane for simplicity. For this ansatz, one can
easily check that equation (\ref{chi4}) has the following solution
\begin{equation}\label{Chi}
\chi=\frac{\mathcal{C}}{a^4}+\frac{4\gamma A}{\mu+4}a^{\mu},
\end{equation}
where $\mathcal{C}$ is an integration constant usually referred to
the dark radiation term. In a similar way, inserting ${T}^{0}_{\ 5}$
into equation (\ref{T2}), we get
\begin{equation}
\rho=\frac{\rho_0}{ a^{3}}-\frac{2A}{\mu+3}a^{\mu},
\end{equation}
where $\rho_{0}$ is the present matter density of the Universe with
equation of state $w=0$. Finally, inserting $\rho$ and $\chi$ into
equation (\ref{fri4}), we obtain the modified Friedmann equation on
the brane
\begin{equation}
H^2\,=\,\frac{8\pi
G_N}{3}\left(\rho_m-\frac{2A}{(\mu+3)(\mu+4)}a^{\mu}\right)-\frac{K}{a^2}\,,{\label{Fried}}
\end{equation}
where $G_N=3\gamma/4\pi$ is the $4$D Newtonian constant,
$\rho_m=\rho_{0} a^{-3}$ is matter energy density and we have
neglected the dark radiation term $\sim a^{-4}$, namely
${\mathcal{C}}= 0$, because we are more interested in the prob of
late time era. Using the value of present critical density,
\begin{equation}
\rho_{c}=\frac{3H_{0}^2}{8\pi G_N},
\end{equation}
the effective Friedmann equation in terms of dimensionless
quantities and redshift parameter $1+z=a^{-1}$ can be written as
 \begin{equation}\label{hub}
H^2=H_0^2[\Omega_m (1+z)^{3}-\Omega_A (1+z)^{-\mu}-(\Omega_{{\rm
tot}}-1)(1+z)^{2}],
\end{equation}
where
\begin{eqnarray}
\Omega_{m}&=&\frac{\rho_{0}}{\rho_{c}}, \hspace{0.5cm}    \Omega_{K}=\frac{3K}{8\pi G_N\rho_{c}}, \\
\Omega_{A}&=&\frac{2A}{\rho_{c}(\mu+3)(\mu+4)},\\
\Omega_{{\rm tot}}&=&\Omega_m-\Omega_{A}=1+\Omega_K.
\end{eqnarray}
As one can see from equation (\ref{hub}), the free parameters of
this model are very similar to those of $\Lambda$CDM models, but
this is quite accidental and is due to our specific ansatz for the
energy exchange term ${T}^{0}_{\ 5}$. Indeed, the energy exchange
term $\Omega_A$ in equation (\ref{hub}) which behaves such as
cosmological constant term in $\Lambda$CDM models is originated from
the bulk matter content (see equation (\ref{bulk})). This is
completely different from the origin of the corresponding term in
the $\Lambda$CDM models. In order to get the late time acceleration
expansion profile for the Universe, this term plays a crucial role
here. Therefore the braneworld model with energy exchange between
the brane and bulk, gives a very useful framework for comparing the
$\Lambda$CDM general relativistic cosmology to a modified gravity
alternative.

\begin{figure}
\begin{center}
\includegraphics[width=\columnwidth]{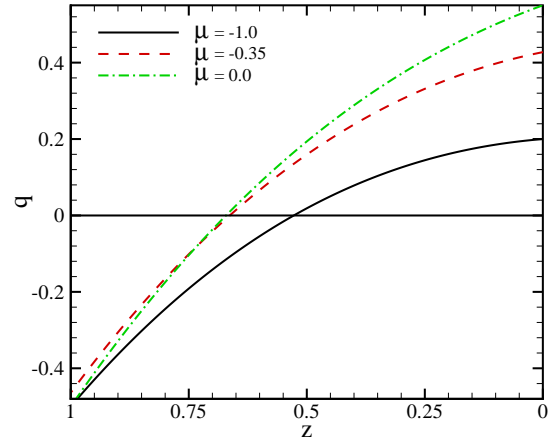}
\caption{Acceleration parameter ($q=\ddot{a}/aH_0^2$) in the
braneworld model as a function of redshift for various values of
$\mu$. Here we chose $\Omega_K=0.0$, $\Omega_m=0.30$ and
$\Omega_A=-0.70$.} \label{qbrane}
\end{center}
\end{figure}

\begin{figure}
\begin{center}
\includegraphics[width=\columnwidth]{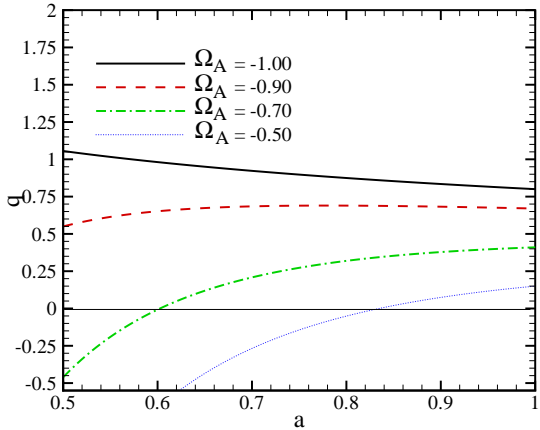}
\includegraphics[width=\columnwidth]{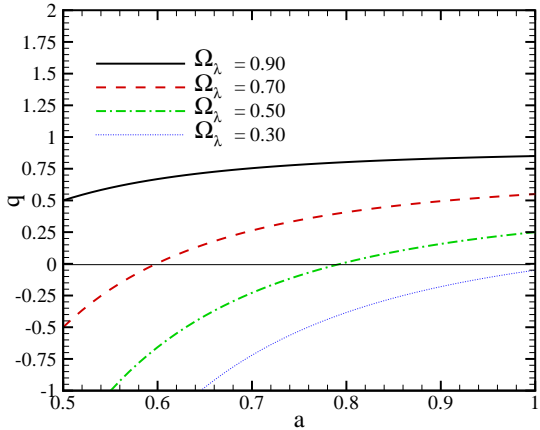}
\caption{Upper panel shows acceleration parameter
$(q=\ddot{a}/{H_0^2a})$ in the braneworld model as a function of
scale factor for various values of $\Omega_A$ and $\mu=-0.40$. Lower
panel corresponds to the same function for the flat $\Lambda$CDM. We
chose the flat Universe.} \label{acc}
\end{center}
\end{figure}

To see how our model can exhibit acceleration expansion of our
Universe, we study the behavior of the acceleration parameter. One
can easily show that the acceleration parameter in this model can be
written as
\begin{equation}
q\equiv \frac{1}{{H^2_0}}\frac{{\ddot
a}}{a}=-\frac{1}{2}\left[\Omega_m(1+z)^3+\Omega_A(\mu+2)(1+z)^{-\mu}\right].
\end{equation}\label{q}
As it can be seen in figure \ref{qbrane}, increasing $\mu$ causes
the Universe to accelerate earlier. In figure \ref{acc} we compare
this model and $\Lambda$CDM just according to the acceleration
parameter. Increasing (decreasing) the value of $\Omega_A$
($\Omega_{\lambda}$ in $\Lambda$CDM model) causes the Universe to
enter accelerating epoch earlier. As we will see in the following
section using the best fit values for model parameters, acceleration
parameter in the present time at $1\sigma$ confidence level is
$q(z=0)=0.42_{-0.19}^{+0.21}$ while for $\Lambda$CDM model is
$q(z=0)=0.59_{-0.02}^{+0.02}$.

\begin{figure}
\begin{center}
\includegraphics[width=\columnwidth]{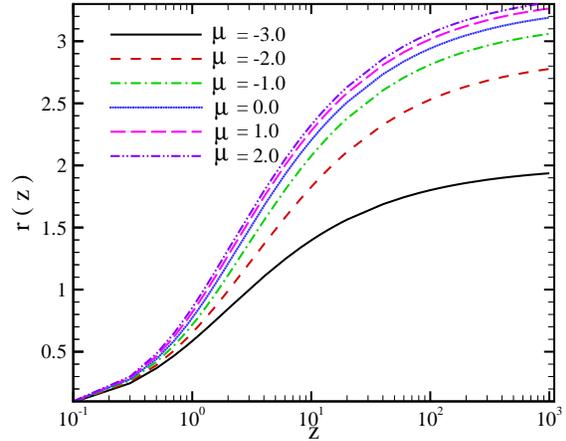}
\caption{Comoving distance, $r(z;\Omega_m,\Omega_{A},\mu)$ (in unit
of $c/H_0$) as a function of redshift for various values of $\mu$.
We fixed $\Omega_K= 0.0$, $\Omega_m=0.3$ and $\Omega_A=-0.7$.}
\label{fig:rz}
\end{center}
\end{figure}


\begin{figure}
\begin{center}
\includegraphics[width=\columnwidth]{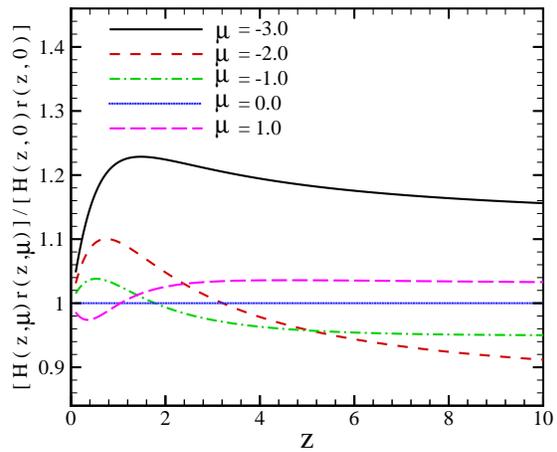}
\caption{Alcock-Paczynski test comparing $\Delta z/{\Delta \theta}$
as a function of redshift for five different $\mu$ normalized to the
case with $\Omega_m=0.30$ and $\Omega_A=-0.70$ and $\mu=0.0$ (flat
Universe $\Omega_K=0$ and $\Lambda$CDM ).} \label{fig:hr}
\end{center}
\end{figure}


\begin{figure}
\begin{center}
\includegraphics[width=\columnwidth]{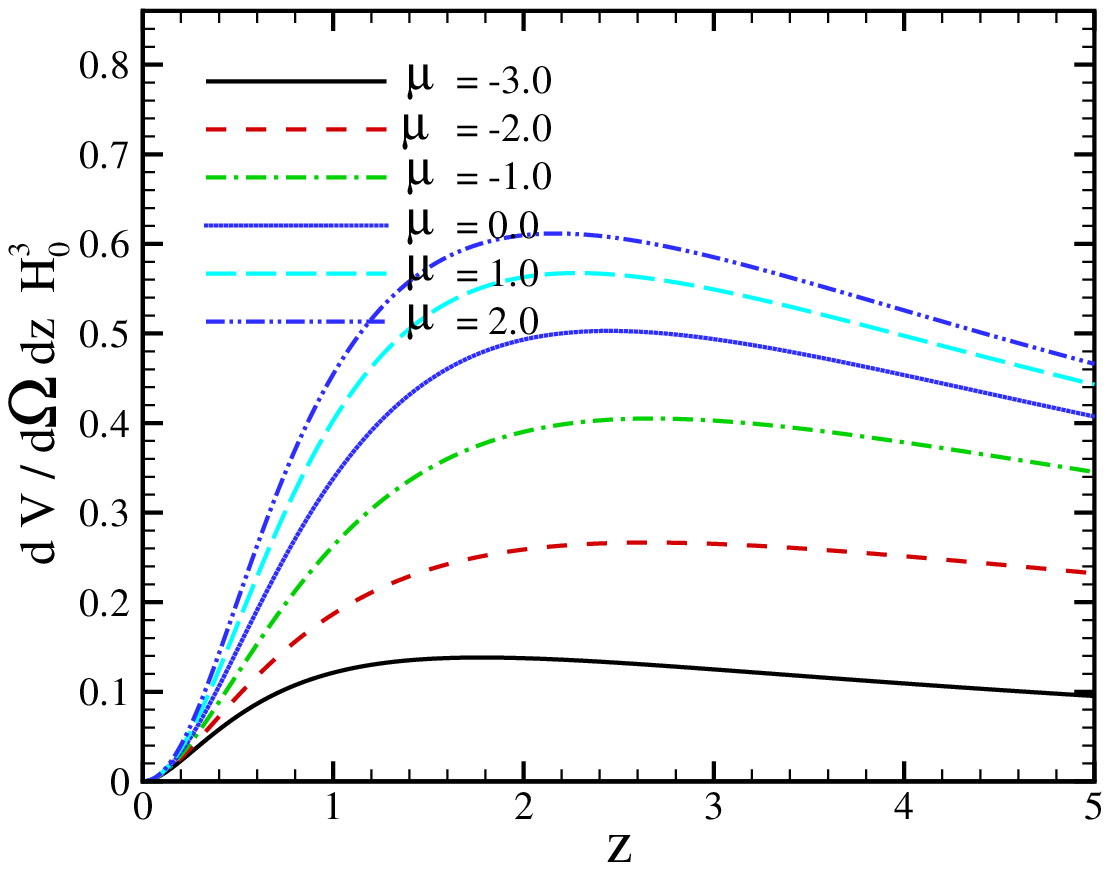}
\caption{The comoving volume element in terms of redshift for
various $\mu$ exponent. Increasing $\mu$ shifts the position of the
maximum value of volume element to higher redshifts. We fixed
$\Omega_K=0.0$.} \label{fig:v}
\end{center}
\end{figure}

Now an interesting question that arises is: \textit{can this model
predict dynamics of the Universe}? In other words, \textit{ For what
values of the free parameters, theoretical model is consistent with
the observational tests? }

In the forthcoming sections we will see what constraints to the
described model are set by recent observations. As a matter of fact
we  examine the free parameters of model more carefully. Indeed we
let the parameters scan their phase space and using likelihood
statistics, the best fit values which maximize likelihood function
will be retrieved.
\section{Geometrical Effects of Braneworld Model}

The cosmological observations are mainly affected by the background
dynamics of the Universe. So before starting some main observational
tests to explore braneworld cosmology we investigate how the free
parameters of this model alter the background dynamics by using the
measurable quantities introduced in this section. We believe they
give deep insight throughout this model. For this purpose, we study
the effect of the braneworld model on the geometrical parameters of
the Universe all together.

\subsection{comoving distance}
The radial comoving distance is one of the basic parameters of
cosmology. For an object with the redshift of $z$, using the null
geodesics in the FRW metric, the comoving distance is obtained as:
\begin{eqnarray} r(z;\Omega_m,\Omega_A,\mu) &=& {1 \over
H_0\sqrt{|\Omega_K|}}\, {\cal F} \left( \sqrt{|\Omega_K|}\int_0^z\,
{dz' \over H(z')/H_0} \right), \label{comoving}\nonumber\\
 \end{eqnarray}
where
 \begin{eqnarray}
 {\cal F}(x) &\equiv & (x,\sin x, \sinh
x)~{\rm for}~K=(0,1,-1)\,,
 \end{eqnarray}
and $H(z;\Omega_m,\Omega_A,\mu)$ is given by equation (\ref{hub}).

By numerical integration of equation (\ref{comoving}), the comoving
distance in terms of redshift for different values of $\mu$ is shown
in figure \ref{fig:rz}. Increasing the $\mu$ results in a longer
comoving distance. According to this behavior by fine tuning the
value of $\mu$ in addition to $\Omega_A$ and $\Omega_m$, one can
expect to explain the  observational results given by supernova as a
standard candle to measure distance in the observational cosmology.

\subsection{Angular Size}
The apparent angular size of an object located at the cosmological
distance is another important parameter that can be affected by the
cosmological model during the history of the Universe. An object
with the physical size of $D$ is related to the apparent angular
size of $\theta$ by:
\begin{equation}
D=d_A \theta, \label{as}
\end{equation}
where $d_A=r(z;\Omega_m,\Omega_A,\mu)/(1+z)$ is the angular diameter
distance. The main applications of equation (\ref{as}) is on the
measurement of the apparent angular size of acoustic peak on CMB and
baryonic acoustic peak at the high and low redshifts, respectively.
By measuring the angular size of an object in different redshifts
(the so-called Alcock-Paczynski test) it is possible to probe the
validity of braneworld  model \citep{alc79}. The variation of
apparent angular size $\Delta\theta$ in terms of $\Delta z$ is given
by:
\begin{equation}
{\Delta z\over \Delta \theta} =
H(z;\Omega_m,\Omega_A,\mu)r(z;\Omega_m,\Omega_A,\mu). \label{alpa}
\end{equation}
Figure \ref{fig:hr} shows  $\Delta z/ \Delta \theta$  in terms of
redshift, normalized to the case with $\Omega_{m}=0.30$,
$\Omega_A=-0.70$ and $\mu=0.0$ (flat Universe, $\Omega_K=0.0$). The
advantage of Alcock-Paczynski test is that it is independent of
standard candles and a standard ruler such as the size of baryonic
acoustic peak can be used to constrain the braneworld model.
\subsection{Comoving Volume Element}
The comoving volume element is another geometrical parameter which
is used in number-count tests such as lensed quasars, galaxies, or
clusters of galaxies. The comoving volume element in terms of
comoving distance and Hubble parameter is given by:
\begin{eqnarray}
f(z;\Omega_m,\Omega_A,\mu)& \equiv& {dV\over dz d\Omega}\nonumber\\
& =& r^2(z;\Omega_m,\Omega_A,\mu)/H(z;\Omega_m,\Omega_A,\mu).
\end{eqnarray}
According to figure \ref{fig:v}, the comoving volume element becomes
large for larger value of $\mu$ in the flat Universe.


\section{Observational constraints on the model using background evolution of the Universe}
In this section, at the beginning, we examine braneworld model by
SNIa Gold sample and supernova Legacy Survey data. Then to make the
model parameter intervals more confined, we will combine
observational results of SNIa distance modules with power spectrum
of cosmic microwave background radiation and baryon acoustic
oscillation measured by Sloan Digital Sky survey. Table 1 shows
different priors of the model parameters used in the likelihood
analysis.

\begin{table}
\begin{center}
  \begin{tabular}{rccr}
            \hline
            \noalign{\smallskip}
{\rm Parameter}& Prior & \\  \hline
  $\Omega_{{\rm tot}}=\Omega_m-\Omega_{A}$ & $-$ & {\rm
Free}\\
 $\Omega_m$& $[0.00,1.00]$& {\rm Top hat}\\
 $\Omega_bh^2$&$0.020\pm0.005$&{\rm Top hat (BBN)}\\

$\Omega_A$&$[-3.00,1.00]$&{\rm Top hat}\\
 $h$&$-$&{\rm Free }\\
 $\mu$&$-$&{\rm Free}\\\hline
  \end{tabular}
\caption{ Priors of the parameter space, used in the likelihood
analysis \citep{bbn,hst,zang}. }
\end{center}

\label{table1}
\end{table}

 \subsection{Supernova Type Ia: Gold and SNLS Samples } \label{cobs}
The Supernova Type Ia experiments provided the main evidence of the
existence of dark energy. Since 1995 two teams of the {\it High-Z
Supernova Search} and the {\it Supernova Cosmology Project} have
discovered several type Ia supernovas at the high redshifts
\citep{per99,Schmidt}. Recently, Riess et al.(2004) have announced
the discovery of $16$ type Ia supernova with the Hubble Space
Telescope. They determined the luminosity distance of these
supernovas and with the previously reported algorithms, obtained a
uniform $157$ Gold sample of type Ia supernovas
\citep{R04,Tonry,bar04}. Recently a new data set of Gold sample with
smaller systematic error containing $156$ Supernova Ia has been
released \citep{new}. In this work we use this data set as new Gold
sample SNIa.

\begin{figure}
\begin{center}
\includegraphics[width=\columnwidth]{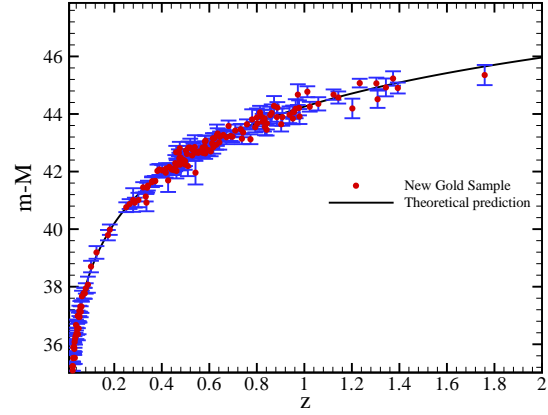}
\caption{Distance modulus of the SNIa new Gold sample in terms of
redshift. Solid line shows the best fit values with the
corresponding parameters of $H_0=63.66$,
$\Omega_m=0.51^{+0.10}_{-0.30}$, $\Omega_A=-0.75_{-1.41}^{+0.32}$
and $\mu=0.76_{-1.95}^{+7.24}$
 in $1 \sigma$ level of
confidence with $\chi^2_{min}/N_{d.o.f} =0.92$ for brane world
model.} \label{modul1}
\end{center}
\end{figure}

\begin{figure}
\begin{center}
\includegraphics[width=\columnwidth]{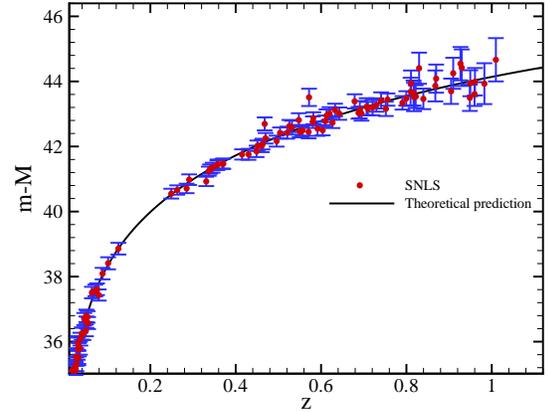}
\caption{Distance modulus of the SNLS supernova data in terms of
redshift. Solid line shows the best fit values with the
corresponding parameters of $H_0=69.38$,
$\Omega_m=0.06^{+0.44}_{-0.06}$, $\Omega_{A}=-1.84_{-0.59}^{+1.58}$
 and $\mu=-1.34^{+7.34}_{-0.10}$  in $1 \sigma$ level of
confidence with $\chi^2_{min}/N_{d.o.f} =0.87$ for brane world
model.} \label{modul2}
\end{center}
\end{figure}

\begin{figure}
\begin{center}
\includegraphics[width=\columnwidth]{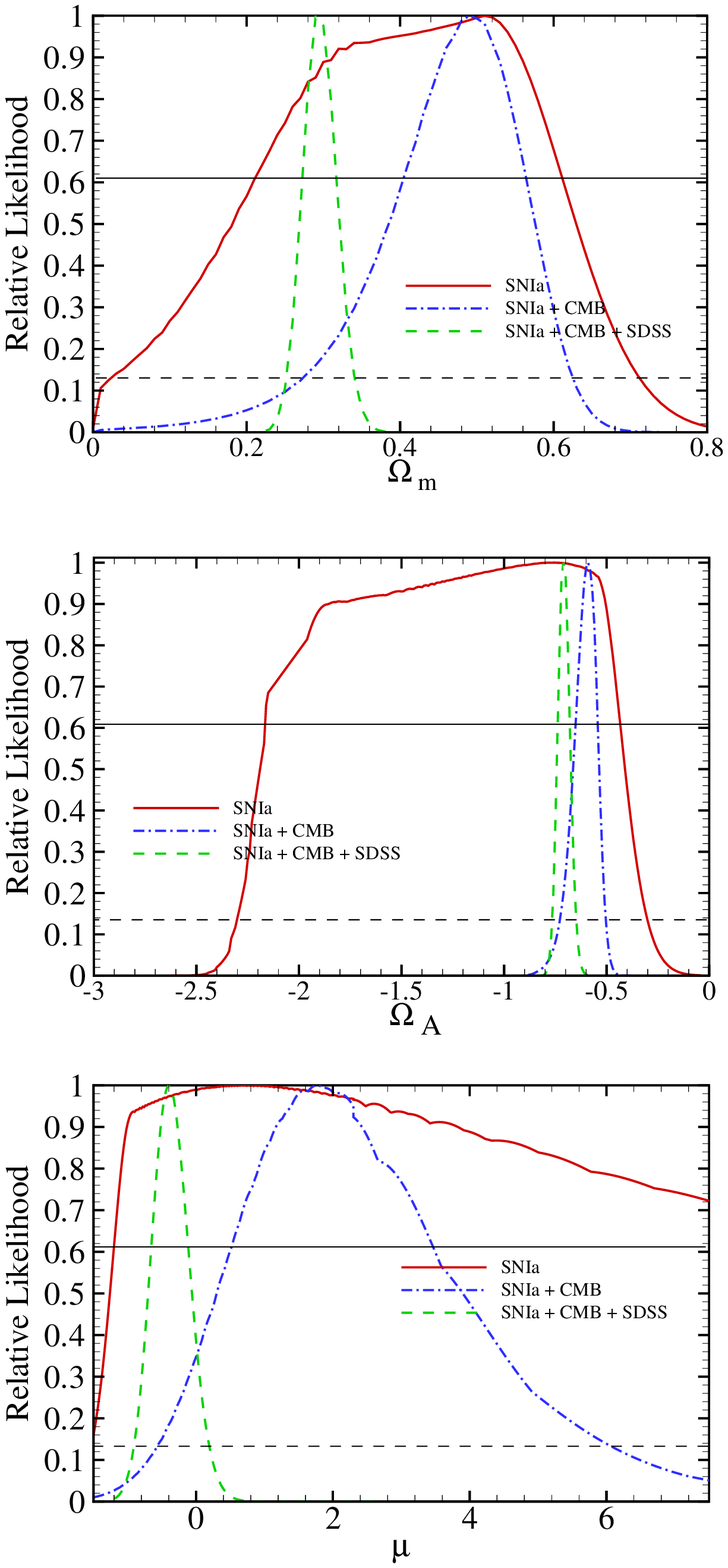}
\caption{Marginalized likelihood functions of three parameters of
model ($\Omega_m$, $\Omega_A$ and $\mu$). The solid line corresponds
to the likelihood function of fitting the model with SNIa data (new
Gold sample), the dashdot line with the joint SNIa$+$CMB data and
dashed line corresponds to SNIa$+$CMB$+$SDSS. The intersections of
the curves with the horizontal solid and dashed lines give the
bounds with $1\sigma$ and $2\sigma$ level of confidence
respectively.} \label{like1}
\end{center}
\end{figure}

\begin{figure}
\begin{center}
\includegraphics[width=\columnwidth]{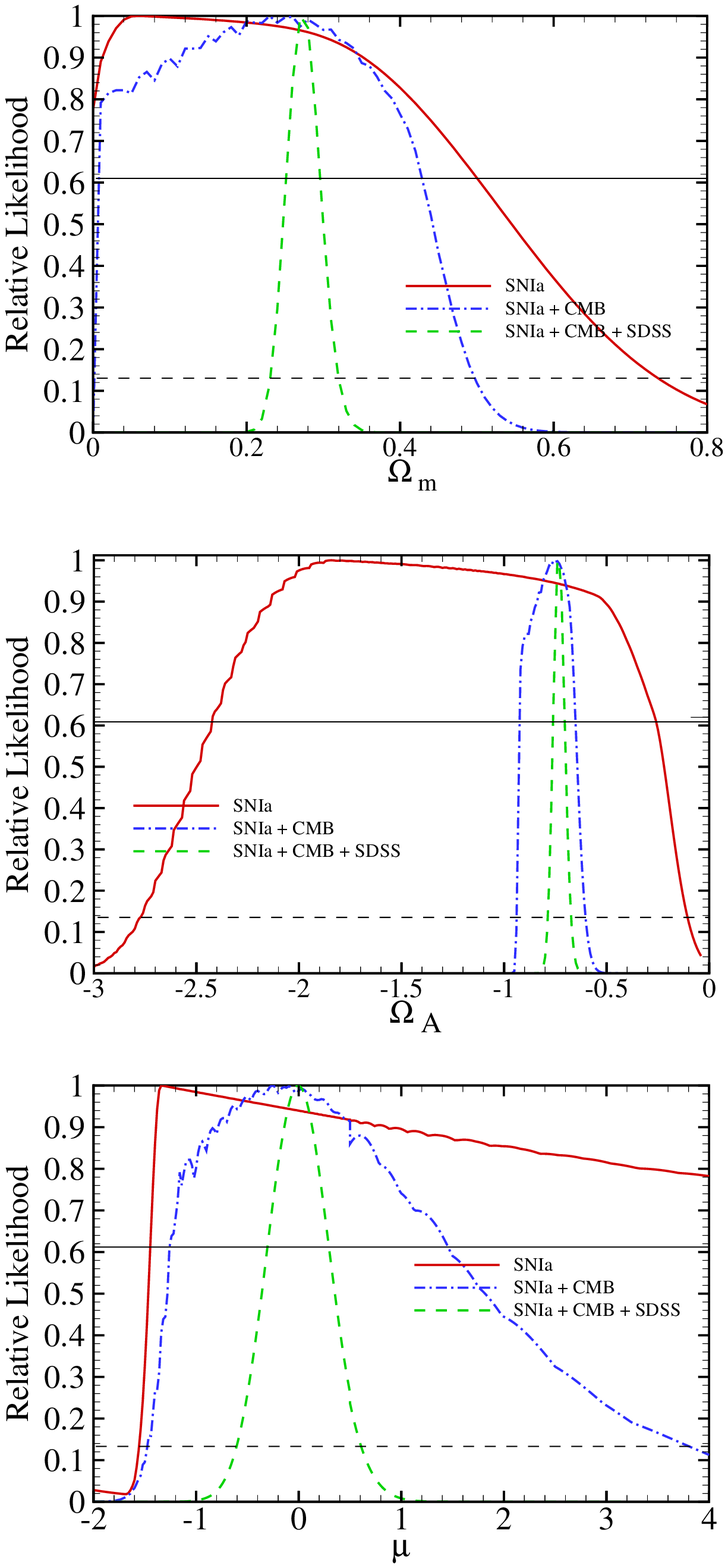}
\caption{Marginalized likelihood functions of three parameters of a
braneworld model ($\Omega_m$, $\Omega_A$ and $\mu$). The solid line
corresponds to the likelihood function of fitting the model with
SNIa data (SNLS), the dashdot line with the joint SNIa$+$CMB data
and dashed line corresponds to SNIa$+$CMB$+$SDSS. The intersections
of the curves with the horizontal solid and dashed lines give the
bounds with $1\sigma$ and $2\sigma$ level of confidence
respectively.} \label{like2}
\end{center}
\end{figure}

More recently, the SNLS collaboration released the first year data
of its planned five-year Supernova Legacy Survey\citep{astier05}. An
important aspect to be emphasized on the SNLS data is that they seem
to be in a better agreement with WMAP results than the Gold sample
\citep{pad06}. We compare the predictions of the braneworld model
for apparent magnitude with new SNIa Gold sample and SNLS data set.
The observations of supernova measure essentially the apparent
magnitude $m$ including reddening, K correction, etc, which are
related to the (dimensionless) luminosity distance, $D_L$, of an
object at redshift $z$ through
\begin{equation}
m={\mathcal{M}}+5\log{D_{L}(z;\Omega_m,\Omega_A,\mu)} \label{m},
\end{equation} where
\begin{eqnarray}
\label{luminosity} D_L (z;\Omega_m,\Omega_A,\mu) &=&{(1+z) \over
\sqrt{|\Omega_K|}}\, {\cal F} \left( \sqrt{|\Omega_K|}\int_0^z\,
{dz'H_0\over H(z')} \right).
\end{eqnarray}
Also
\begin{eqnarray}
\label{m1}\mathcal{M} &=& M+5\log{\left(\frac{c/H_0}{1\quad
Mpc}\right)}+25,
\end{eqnarray}
where $M$ is the absolute magnitude. The distance modulus, $\Re$, is
defined as

\begin{eqnarray}
\Re\equiv
m-M&=&5\log{D_{L}(z;\Omega_m,\Omega_A,\mu)}\nonumber\\&&\quad
+5\log{\left(\frac{c/H_0}{1\quad Mpc}\right)}+25, \label{eq:mMr}
\end{eqnarray}
or

\begin{eqnarray}
\Re&=&5\log{D_{L}(z;\Omega_m,\Omega_A,\mu)}+\bar{M}.
\end{eqnarray}
In order to compare the theoretical results with the observational
data, we must compute the distance modulus, as given by equation
(\ref{eq:mMr}). For this purpose, the first step is to compute the
quality of the fitting through the least squared fitting quantity
$\chi^2$ defined by:
\begin{eqnarray}\label{chi_sn}
&&\chi^2(\bar{M},\Omega_m,\Omega_A,\mu)=\nonumber\\
&&\sum_{i}\frac{[\Re_{{\rm obs}}(z_i)-\Re_{{\rm
th}}(z_i;\Omega_m,\Omega_A,\mu,\bar{M})]^2}{\sigma_i^2},
\end{eqnarray}
where $\sigma_i$ is the observational uncertainty in the distance
modulus. To constrain the parameters of model, we use the Likelihood
statistical analysis:
\begin{eqnarray}
{\cal
L}(\bar{M},\Omega_m,\Omega_A,\mu)={\mathcal{N}}e^{-\chi^2(\bar{M},\Omega_m,\Omega_A,\mu)/2},
\end{eqnarray}
where ${\mathcal{N}}$ is a normalization factor. The parameter
$\bar{M}$ is a nuisance parameter and should be marginalized
(integrated out) leading to a new $\bar{\chi}^2$ defined as:
\begin{eqnarray}\label{mar2}
\bar{\chi}^2=-2\ln\int_{-\infty}^{+\infty}{\cal
L}(\bar{M},\Omega_m,\Omega_A,\mu)d\bar{M}.
\end{eqnarray}
Using equations (\ref{chi_sn}) and (\ref{mar2}), we find
\begin{eqnarray}\label{mar3}
\bar{\chi}^2(\Omega_m,\Omega_A,\mu)&=&\chi^2(\bar{M}=0,\Omega_m,\Omega_A,\mu)\nonumber\\&&
-\frac{B(\Omega_m,\Omega_A,\mu)^2}{C}+\ln(C/2\pi),
\end{eqnarray}
where
\begin{eqnarray}\label{mar4}
&&B(\Omega_m,\Omega_A,\mu)=\nonumber\\
&&\sum_{i}\frac{[\Re_{{\rm obs}}(z_i)-\Re_{{\rm
th}}(z_i;\Omega_m,\Omega_A,\mu,\bar{M}=0)]}{\sigma_i^2},
\end{eqnarray}
and
\begin{eqnarray}\label{mar5}
C=\sum_{i}\frac{1}{\sigma_i^2}.
\end{eqnarray}
Equivalent to marginalization is the minimization with respect to
$\bar{M}$. One can show that $\chi^2$ can be expanded in $\bar{M}$
as \citep{Nesseris04}
\begin{eqnarray}\label{mar6}
\chi^2_{\rm
SNIa}(\Omega_m,\Omega_A,\mu)&=&\chi^2(\bar{M}=0,\Omega_m,\Omega_A,\mu)\nonumber\\&&
-2\bar{M}B+\bar{M}^2C,
\end{eqnarray}
which has a minimum for $\bar{M}=B/C$:
\begin{eqnarray}\label{mar7}
\chi^2_{\rm SNIa}(\Omega_m,\Omega_A,\mu)&=&\chi^2(\bar{M}=0,\Omega_m,\Omega_A,\mu)\nonumber\\
&&-\frac{B(\Omega_m,\Omega_A,\mu)^2}{C}.
\end{eqnarray}
Using equation (\ref{mar7}) we can find the best fit values of model
parameters as the values that minimize $\chi^2_{\rm
SNIa}(\Omega_m,\Omega_A,\mu)$. The best fit values for the
parameters of model by using supernova data are
$\Omega_m=0.51^{+0.10}_{-0.30}$, $\Omega_A=-0.75_{-1.41}^{+0.32}$
 and $\mu=0.76^{+7.24}_{-1.95}$ with $\chi^2_{min}/N_{d.o.f} =0.92$
at $1 \sigma$ level of confidence. These values imply that
$\Omega_K=-0.26_{-1.44}^{+0.33}$. The best fit values for the
parameters of model by using SNLS supernova data are
$\Omega_m=0.06^{+0.44}_{-0.06}$, $\Omega_A=-1.84_{-0.59}^{+1.58}$
 and $\mu=-1.34^{+7.34}_{-0.10}$ with $\chi^2_{min}/N_{d.o.f} =0.87$
at $1 \sigma$ level of confidence. The corresponding value of
$\Omega_K$ at $1\sigma$ confidence level is
$\Omega_K=-0.90^{+1.64}_{-0.59}$. Figures \ref{modul1} and
\ref{modul2} show the comparison of the theoretical prediction of
distance modulus by using the best fit values of model parameters
and observational values from new Gold sample and SNLS supernova,
respectively. Figures \ref{like1} and  \ref{like2} show relative
likelihood for free parameters of  brane model.

\subsection{Combined analysis: SNIa$+$CMB$+$SDSS} \label{cmb}

\begin{figure}
\begin{center}
\includegraphics[width=\columnwidth]{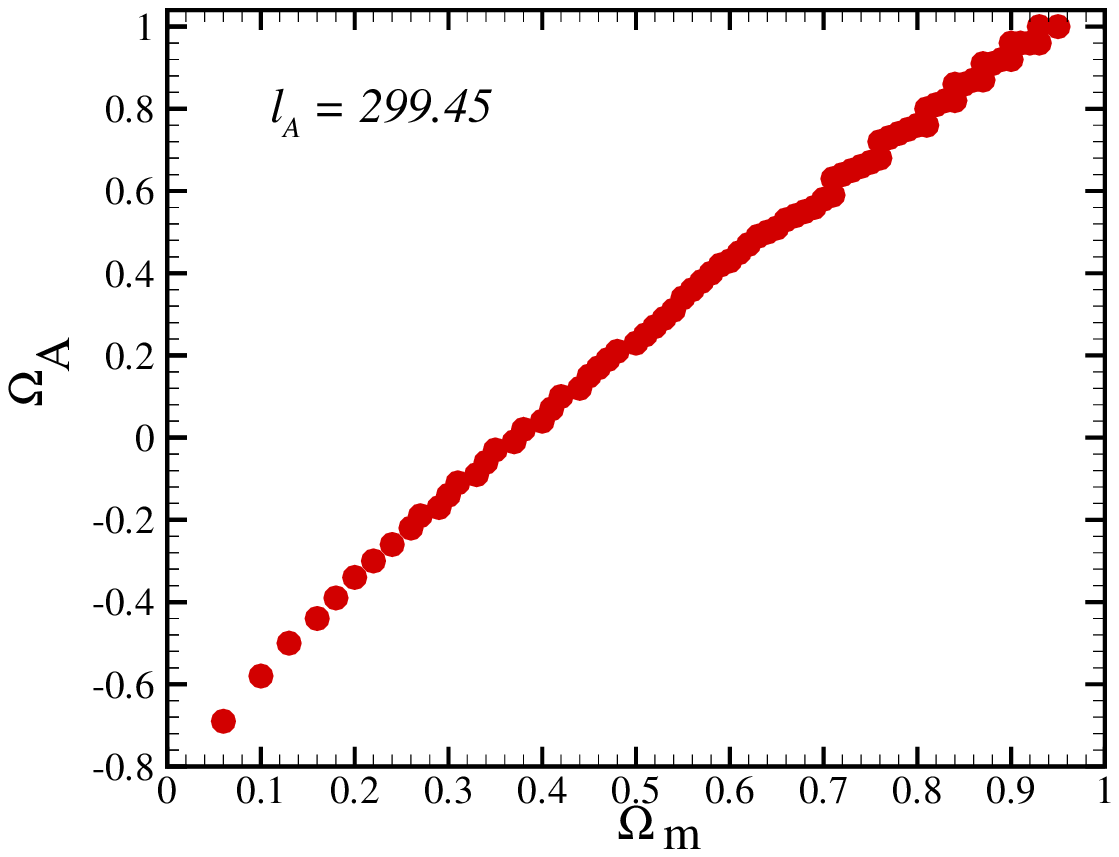}
\includegraphics[width=\columnwidth]{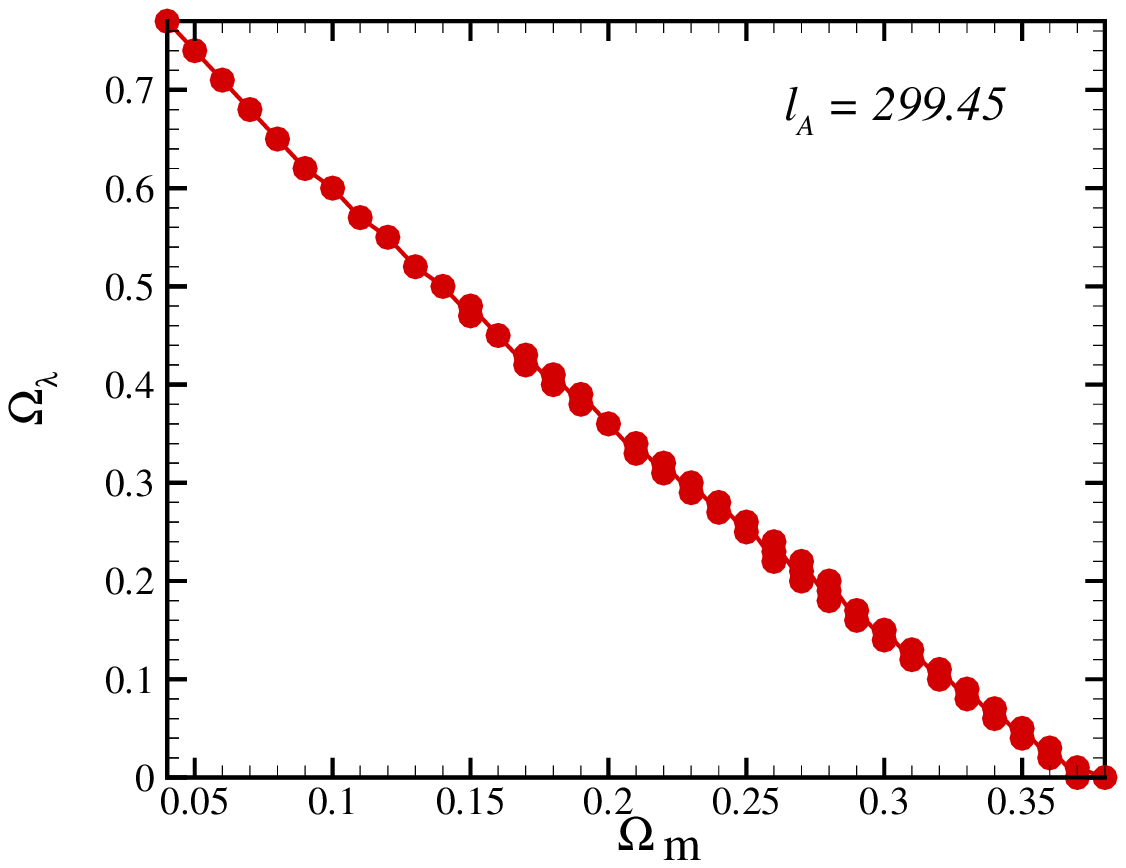}
\caption{Constant acoustic angular scale in the joint space of
$\Omega_m$ and $\Omega_A$ (upper panel). Lower panel shows
dependence of acoustic angular scale on the $\Omega_m$ and
cosmological constant.} \label{l1}
\end{center}
\end{figure}


To obtain more confined acceptable intervals of model free
parameters, now we combine SNIa data (from SNIa new Gold sample and
SNLS) with CMB data from the WMAP and recently observed baryonic
peak from the SDSS. We  also examine the peaks positions of power
spectrum in addition to the common shift parameter.

Before last scattering, the photons and baryons are tightly coupled
by Compton scattering and behave as a fluid. The oscillations of
this fluid, occurring as a result of the balance between the
gravitational interactions and the photon pressure, lead to the
familiar spectrum of peaks and troughs in the averaged temperature
anisotropy spectrum which we measure today. The odd and even peaks
correspond to maximum compression of the fluid and to rarefaction,
respectively \citep{hu96}. In an idealized model of the fluid, there
is an analytic relation for the location of the $m$-th peak: $l_m
\approx ml_A$ \citep{Hu95,Hu} where $l_A$ is the acoustic scale
which may be calculated analytically and depends on both pre- and
post-recombination physics as well as the geometry of the Universe.
The acoustic scale corresponds to the Jeans length of photon-baryon
structures at the last scattering surface some $\sim 379$ Kyr after
the Big Bang \citep{spe03}. The apparent angular size of acoustic
peak can be obtained by dividing the comoving size of sound horizon
at the decoupling epoch $r_s(z_{dec})$ by the comoving distance of
observer to the last scattering surface $r(z_{dec})$
\begin{equation}
\theta_A =\frac{\pi}{l_A}\equiv {{r_s(z_{dec})}\over r(z_{dec}) }.
\label{eq:theta_s}
\end{equation}
The size of sound horizon at the numerator of equation
(\ref{eq:theta_s}) corresponds to the distance that a perturbation
of pressure can travel from the beginning of the Universe up to the
last scattering surface and is given by
 \begin{eqnarray}
&&r_{s}(z_{dec};\Omega_m,\Omega_A,\mu) \nonumber\\
&&= {1 \over H_0\sqrt{|\Omega_K|}}\times {\cal F} \left(
\sqrt{|\Omega_K|}\int_{z_{dec}}^ {\infty} {v_s(z')dz' \over
H(z')/H_0} \right), \label{sh}
 \end{eqnarray}
where $v_s(z)^{-2}=3 + 9/4\times\rho_b(z)/\rho_{rad}(z)$ is the
sound velocity in the unit of speed of light from the big bang up to
the last scattering surface \citep{dor01,Hu95} and the redshift of
the last scattering surface, $z_{dec}$, is given by \citep{Hu95}:
\begin{eqnarray}\label{dec}
z_{dec} &=& 1048\left[ 1 + 0.00124(\omega_b)^{-0.738}\right]\left[
1+g_1(\omega_m)^{g_2}\right],\nonumber \\
g_1 &=& 0.0783(\omega_b)^{-0.238}\left[1+39.5(\omega_b)^{0.763}\right]^{-1},\nonumber \\
g_2 &=& 0.560\left[1 + 21.1(\omega_b)^{1.81}\right]^{-1},
\end{eqnarray}
where $\omega_m\equiv\Omega_mh^2$, $\omega_b\equiv\Omega_bh^2$ and
$\rho_{rad}$ is the radiation density. $\Omega_b$ is relative
baryonic density to the critical density at the present time.
Changing the parameters of the model can change the size of apparent
acoustic peak and subsequently the position of $l_A\equiv
\pi/\theta_A$ in the power spectrum of temperature fluctuations at
the last scattering surface. The simple relation $l_m\approx ml_A$
however does not hold very well for the peaks although it is better
for higher peaks \citep{Hu,doran01}. Driving effects from the decay
of the gravitational potential as well as contributions from the
Doppler shift of the oscillating fluid introduce a shift in the
spectrum. A good parameterization for the location of the peaks and
troughs is given by \citep{Hu,doran01}
\begin{equation}\label{phase shift}
l_m=l_A(m-\phi_m),
\end{equation}
where $\phi_m$ is phase shift determined predominantly by
pre-recombination physics, and are independent of the geometry of
the Universe. The location of acoustic peaks can be determined in
model by equation (\ref{phase shift}) with
$\phi_m(\omega_m,\omega_b)$. Doran et. al. \citep{doran01},
 have recently shown that the first and third phase shifts are
approximately model independent. The values of these shift
parameters have been reported as: $\phi_1(\omega_m,\omega_b)\simeq
0.27$ and $\phi_3(\omega_m,\omega_b)\simeq 0.341$
\citep{Hu,Perci,doran01}. According to the WMAP observations:
$l_1=220.1\pm0.8$ and $l_3=809\pm7$, so the corresponding
observational values of $l_A^{\rm obs}$ read as:
\begin{eqnarray}\label{peak1}
l_A^{\rm obs}|_{l_1}&=&\frac{l_1}{(1-\phi_1)}=299.45\pm2.67,\\
l_A^{\rm obs}|_{l_3}&=&\frac{l_3}{(3-\phi_3)}=304.24\pm2.63,
\end{eqnarray}
their Likelihood statistics are as follows:
 \begin{equation}\label{chil1}
  \chi_{l_1}^2=\frac{\left[l_A^{\rm obs}|_{l_1}-l_A^{\rm
th}|_{l_1}\right]^2}{\sigma_1^2},
\end{equation}
and
\begin{equation}\label{chil3}
\chi_{l_3}^2=\frac{\left[l_A^{\rm obs}|_{l_3}-l_A^{\rm
th}|_{l_3}\right]^2}{\sigma_3^2},
\end{equation}
because of weak dependency of phase shift to the cosmological model
usually another model independent parameter which is so-called shift
parameter ${\cal R}$ from CMB observation as
\begin{equation}
{\cal R}\propto\frac{l_1^{flat}}{l_1},
\end{equation}
are used as another observational test. Where $l_1^{flat}$
corresponds to the flat pure-CDM model with $\Omega_m=1.0$ and the
same $\omega_m$ and $\omega_b$ as the original model.  It is easily
shown that shift parameter is as follows \citep{bond97}
\begin{equation}\label{shift_th}
\label{shift} {\cal R}=
\sqrt{\Omega_m}\frac{D_L(z_{dec},\Omega_m,\Omega_A,\mu)}{(1+z_{dec})}.
\end{equation}
The observational results of CMB experiments correspond to a shift
parameter of ${\cal R}=1.716\pm0.062$ (given by WMAP, CBI, ACBAR)
\citep{spe03,pearson03,Kuo}. One of the advantages of using the
parameter ${\cal R}$ is its independency of Hubble constant. In
order to put constraint on the model from CMB, we compare the
observed shift parameter with that of model using likelihood
statistic as \citep{bond97,bbn,Odman}
\begin{equation}
{\cal{L}}\sim e^{-\chi_{\rm CMB}^2/2},
\end{equation}
where \begin{equation}\label{chi_cmb} \chi_{\rm
CMB}^2=\frac{\left[{\cal R}_{{\rm obs}}-{\cal R}_{{\rm
th}}\right]^2}{\sigma_{\rm CMB}^2},
\end{equation}
where ${\cal R}_{{\rm th}}$ and ${\cal R}_{{\rm obs}}$ are
determined using equation (\ref{shift_th}) and given by observation,
respectively. Figure \ref{l1} shows constant value of $l_A$ in the
joint space parameters $(\Omega_m,\Omega_A)$ and
$(\Omega_m,\Omega_{\lambda})$ for the braneworld and the
$\Lambda$CDM model, respectively. Increasing (decreasing)
$\Omega_{A}$ ($\Omega_{\lambda}$) leads to an increasing in the
value of present matter density to make constant value for $l_A$.
What we found is in agreement with figure \ref{acc}.

Another robust observational approach to investigate cosmological
models is inferring the behavior of the matter power spectrum and
time evolution of gravitational clustering in both linear and
nonlinear regimes. The simplest things to do are solving the
relevant Boltzmann and Einestian equations for various matter
contents in the Universe \citep{dodelson}. Matter power spectrum and
other non-linear effects can be a special tools to discriminate
various models as well as to make more confined acceptable range for
their free parameters (see
\citep{chungpeima99,zhaominma07,tomikoivisito06,german05,german06,seok06,vink00,donc06,seljak98,douglas07,abraham05,reis05,german05,german06,pee,LPChimento2006,LAmendola2000,LAmendola2001,LAmendola2002,LAmendola2003}
for recent reviews).
 The conventional form of matter power spectrum
at the late time is \citep{dodelson}:
\begin{equation}
P(k,a)=2\pi^2\delta_H^2T(k)^2\frac{k^n}{H_0^{3+n}}\left(\frac{D(a)}{D(a=1)}\right)^2
\end{equation}
where $n$ is the spectral index of the primordial adiabatic density
perturbations, $T(k)$ is transfer function determines the evolution
of potential in the radiation-matter equality epoch and in the late
time matter density fluctuations govern by so-called growth
function, $D(a)$. $\delta_H$ is also given by initial condition in
the context of inflation. $k$ is the wave-number of fluctuations in
the Fourier space. Generally, it is well known that if any
interaction between matter and dark energy in addition to the new
kind of matter which are the responsible for the background dynamics
of the Universe to be existed, they alter the matter power spectrum
through the three following effects: The first one is that, the
Hubble parameter in different models causes various dynamics for the
background evolution (e.g. in our model is given by equation
(\ref{hub})) as well as power spectrum. The second effect is due to
the inverse proportional of power spectrum to the matter density for
a fixed potential, so any variation in the present value of matter
density causes the smaller or larger amplitude for power spectrum.
Third effect is related to the fact that in different cosmological
models, the matter-radiation equality epoch, $a_{eq}$ and
subsequently the value of $k_{eq}$ change, so the turning over point
in the power spectrum would be reformed.

Here instead of observational constraint using matter power spectrum
we used the weakly model independent constraint by Baryon acoustic
oscillation and ignore any non-linear effects
\citep{linder1,linder2}. Recently using the observations of large
scale structures from the Sloan Digital Sky Survey (SDSS)
\cite{tegmark104,tegmark204} and Two Degree Field Galaxy Redshift
Survey (2dFGRS) \cite{tegmark02}, one can explore the validity of
cosmological models.

The large scale correlation function measured from $46,748$ {\it
Luminous Red Galaxies} (LRG) spectroscopic sample of the SDSS
includes a clear peak at about  $100$ Mpc $h^{-1}$
\citep{eisenstein05}. This peak was identified with the expanding
spherical wave of baryonic perturbations originating from acoustic
oscillations at recombination. The comoving scale of this shell at
recombination is about $150$Mpc in radius. In other words, this peak
has an excellent match to the predicted shape and the location of
the imprint of the recombination-epoch acoustic oscillation on the
low-redshift clustering matter \citep{eisenstein05}. Recently E. V.
Linder has shown in detail some systematic uncertainties for baryon
acoustic oscillation \citep{linder1,linder2}. Nonlinear mode
coupling related to this fact that ever though baryon acoustic
oscillation is mostly contributed by linear scale, but the influence
of non-linear collapsing has quite broad kernel. In other words, one
might say that baryon acoustic oscillation are $90-99\%$ linear in
comparison to the CMB which is $99.99\%$ linear, so this difference
may affect on various models in different way. Careful works to
constrain on the free parameters of underlying model needs to be
carried out  the effect of non-linear mode coupling in the results
of constraint by SDSS observation. Nevertheless, roughly speaking
regards the acceptance intervals for free parameter cover the real
intervals determined by assuming nonlinearity mode for SDSS
observation \citep{Hu96,Eisen98,Amarzguioui,eisenstein04}.

 A dimensionless and independent of $H_0$ version of SDSS observational
parameter is
\begin{eqnarray} \label{lss1}
{\cal A} &=&D_V(z_{\rm sdss})\frac{\sqrt{\Omega_mH_0^2}}{z_{\rm
sdss}}\nonumber\\
&=&\sqrt{\Omega_m}\left[\frac{H_0D_L^2(z_{\rm
sdss};\Omega_m,\Omega_A,\mu)}{H(z_{\rm
sdss};\Omega_m,\Omega_A,\mu)z_{\rm sdss}^2(1+z_{\rm
sdss})^2}\right]^{1/3},\nonumber\\
\end{eqnarray}
where $D_V(z_{\rm sdss})$ is characteristic distance scale of the
survey with the mean of redshift $z_{\rm sdss}$
\citep{eisenstein05,blak03,ness06}. We use the robust constraint on
the braneworld model using the value of ${\cal A}=0.469\pm0.017$
from the LRG observation at $z_{\rm sdss} = 0.35$
\citep{eisenstein05,eisenstein04}. This observation permits the
addition of one more term in the $\chi^2$ of equations (\ref{mar7})
and (\ref{chi_cmb}) to be minimized with respect to $H(z)$ model
parameters. This term is
\begin{equation}
\chi_{\rm SDSS}^2=\frac{\left[{\cal A}_{\rm obs}-{\cal A}_{\rm
th}\right]^2}{\sigma_{\rm SDSS}^2}.
\end{equation}
This is the third observational constraint for our analysis.

\begin{figure}
\begin{center}
\includegraphics[width=\columnwidth]{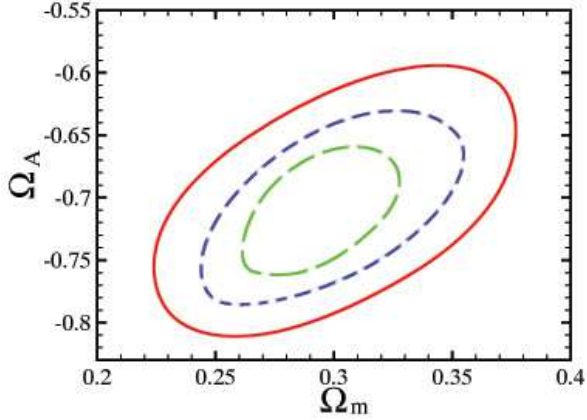}
\caption{Joint confidence intervals of $\Omega_m$ and $\Omega_A$,
fitted with SNIa new Gold sample$+$CMB$+$SDSS. Solid line, dashed
line and long dashed line correspond to $3\sigma$, $2\sigma$ and
$1\sigma$ level of confidence, respectively.} \label{jlike1}
\end{center}
\end{figure}

\begin{figure}
\begin{center}
\includegraphics[width=\columnwidth]{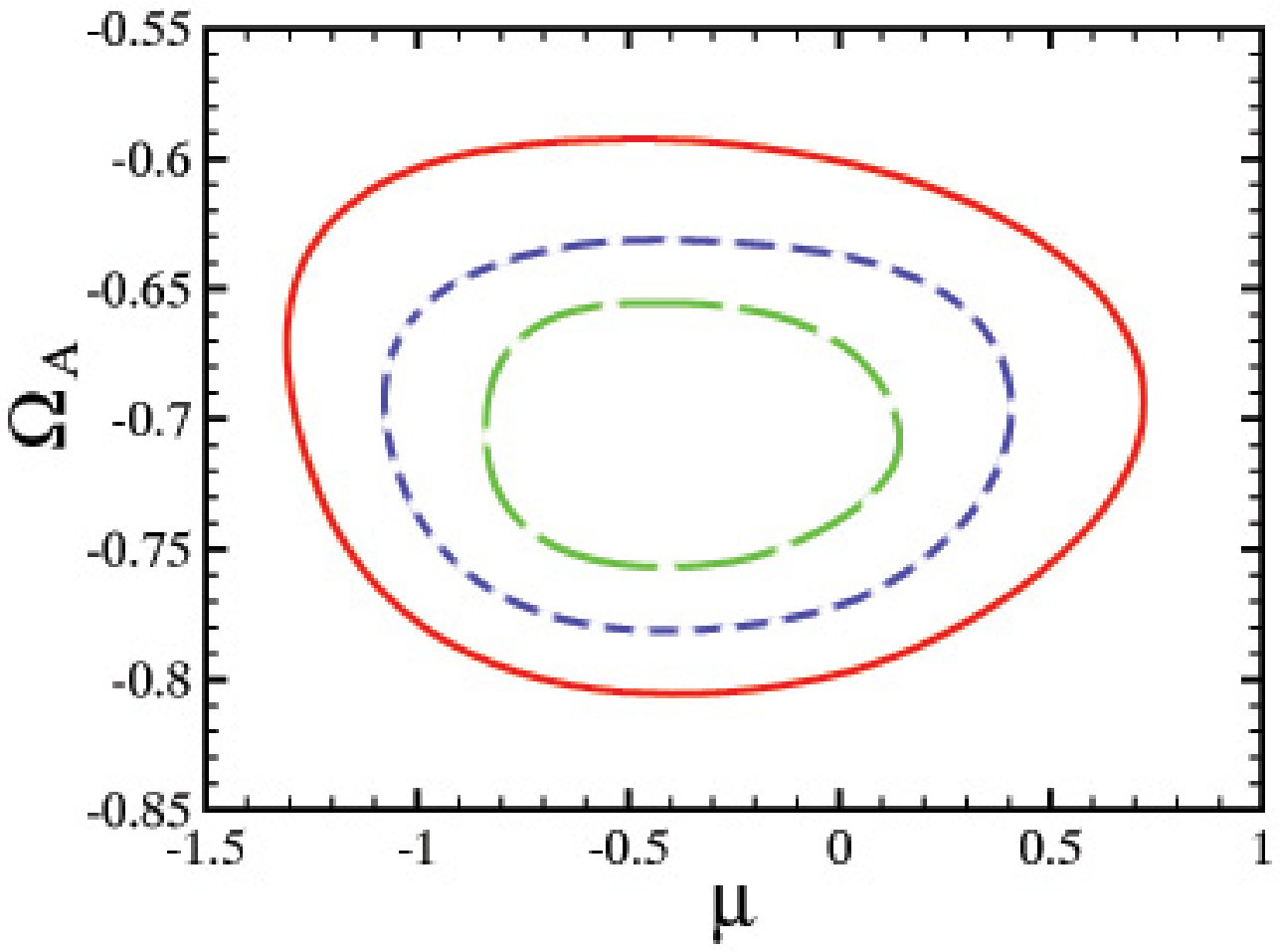}
\caption{Joint confidence intervals of $\mu$ and $\Omega_A$, fitted
with SNIa new Gold sample$+$CMB$+$SDSS. Solid line, dashed line and
long dashed line correspond to $3\sigma$, $2\sigma$ and $1\sigma$
level of confidence, respectively.} \label{jlike2}
\end{center}
\end{figure}

\begin{figure}
\begin{center}
\includegraphics[width=\columnwidth]{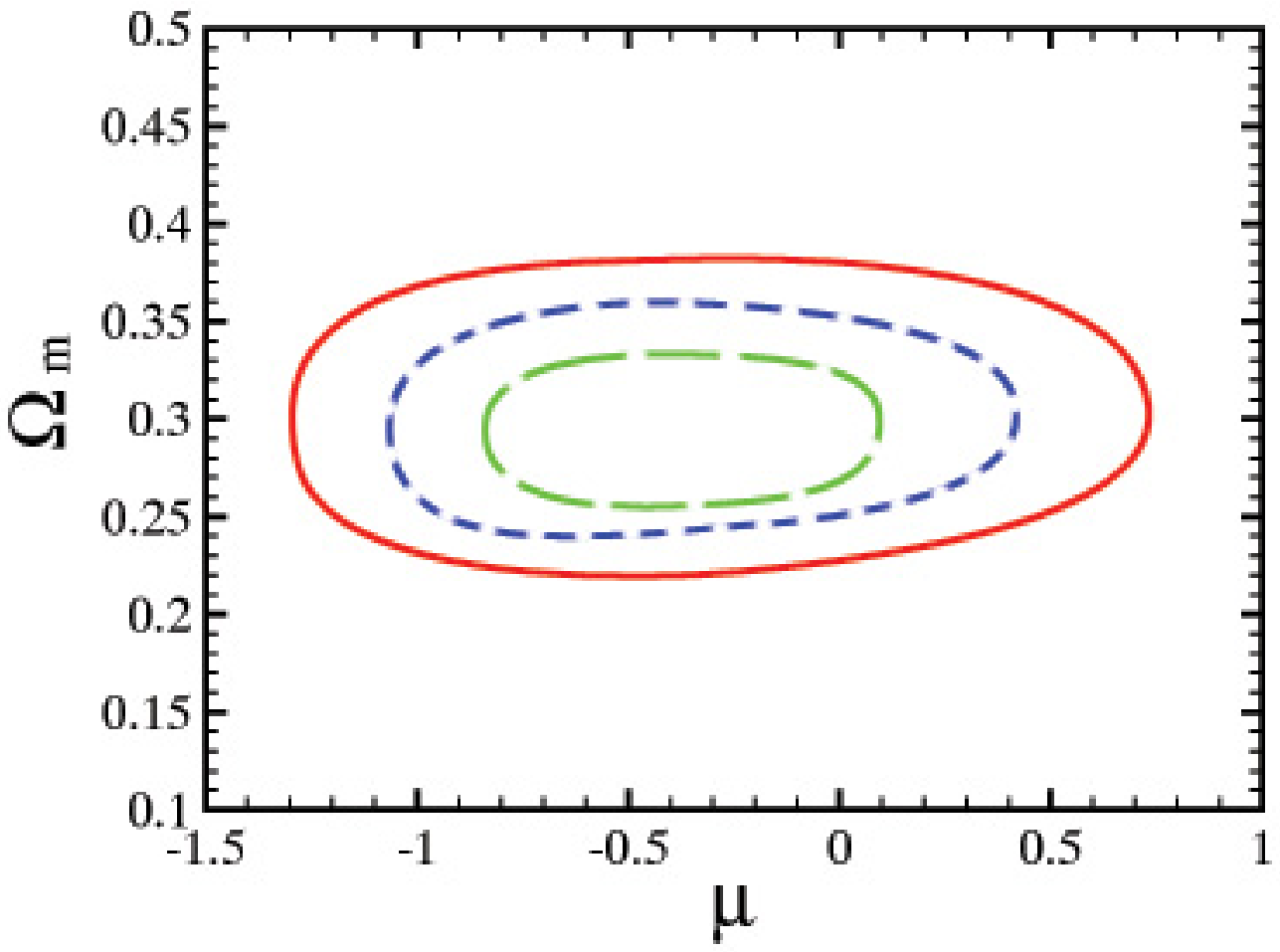}
\caption{Joint confidence intervals of  $\mu$ and $\Omega_m$, fitted
with SNIa new Gold sample$+$CMB$+$SDSS. Solid line, dashed line and
long dashed line correspond to $3\sigma$, $2\sigma$ and $1\sigma$
level of confidence, respectively.} \label{jlike3}
\end{center}
\end{figure}

\begin{figure}
\begin{center}
\includegraphics[width=\columnwidth]{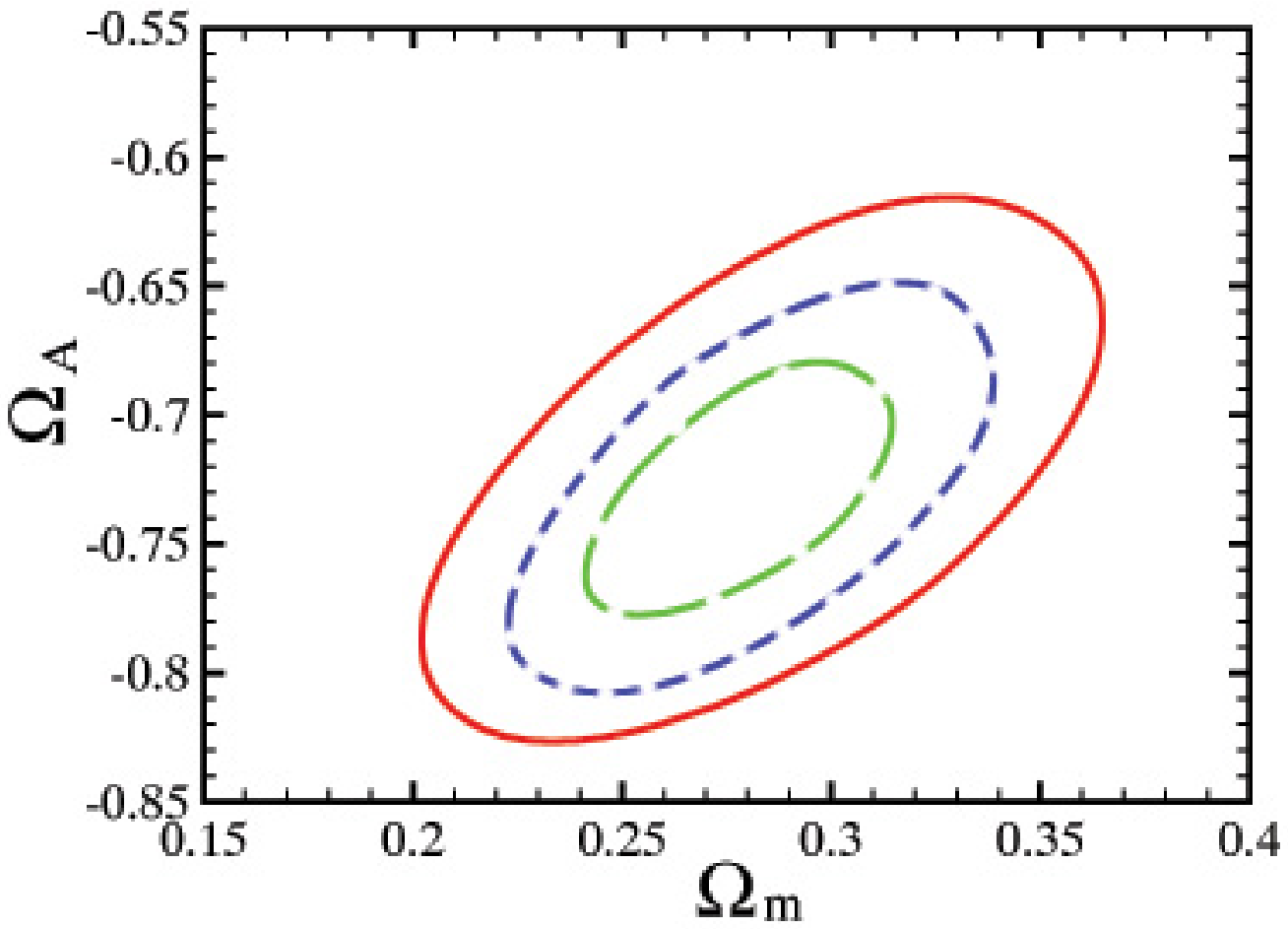}
\caption{Joint confidence intervals of $\Omega_m$ and $\Omega_A$,
fitted with SNIa SNLS$+$CMB$+$SDSS. Solid line, dashed line and long
dashed line correspond to $3\sigma$, $2\sigma$ and $1\sigma$ level
of confidence, respectively.} \label{jlike4}
\end{center}
\end{figure}

\begin{figure}
\begin{center}
\includegraphics[width=\columnwidth]{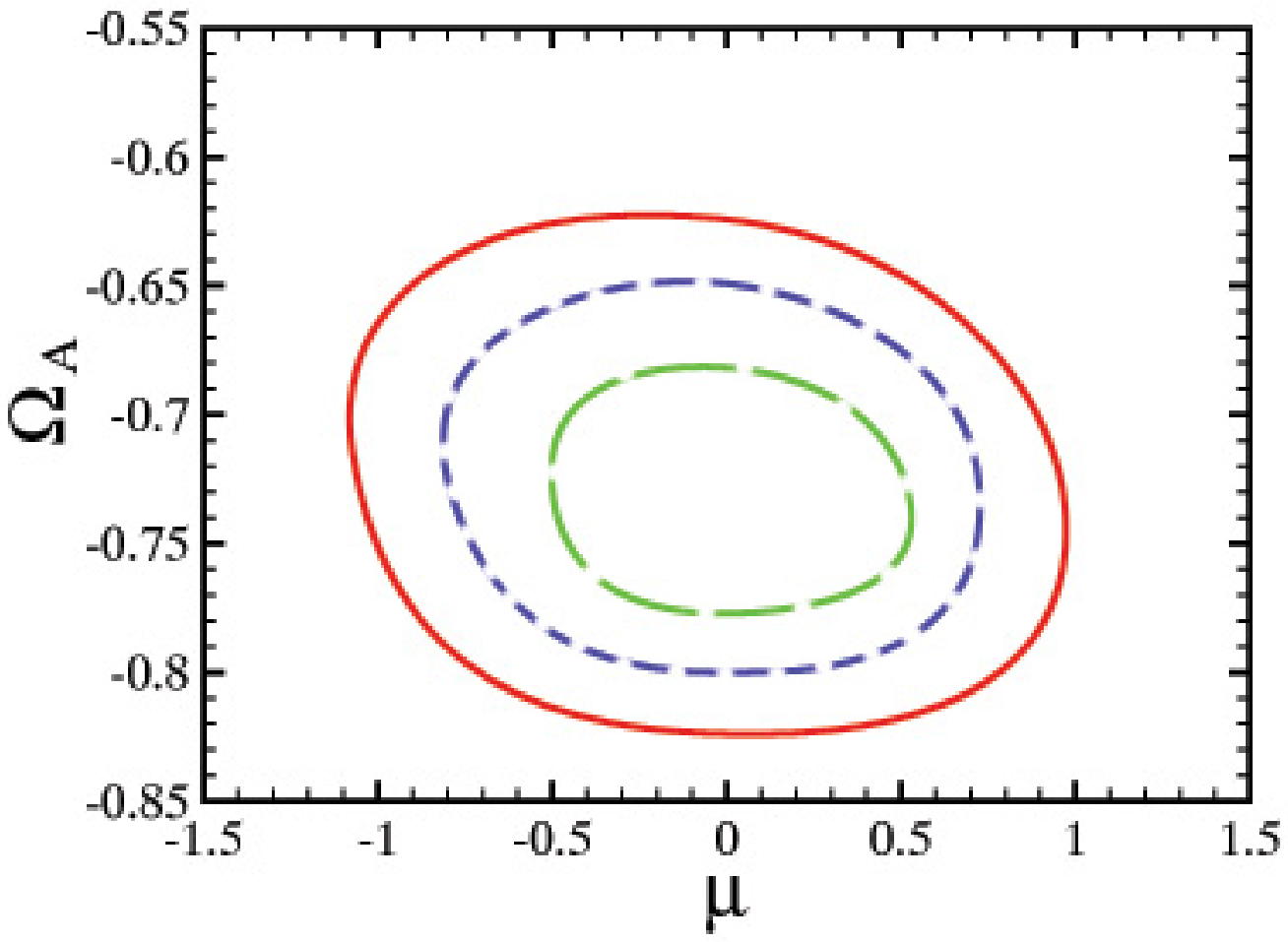}
\caption{Joint confidence intervals of $\mu$ and $\Omega_A$, fitted
with SNIa SNLS$+$CMB$+$SDSS. Solid line, dashed line and long dashed
line correspond to $3\sigma$, $2\sigma$ and $1\sigma$ level of
confidence, respectively.} \label{jlike5}
\end{center}
\end{figure}

\begin{figure}
\begin{center}
\includegraphics[width=\columnwidth]{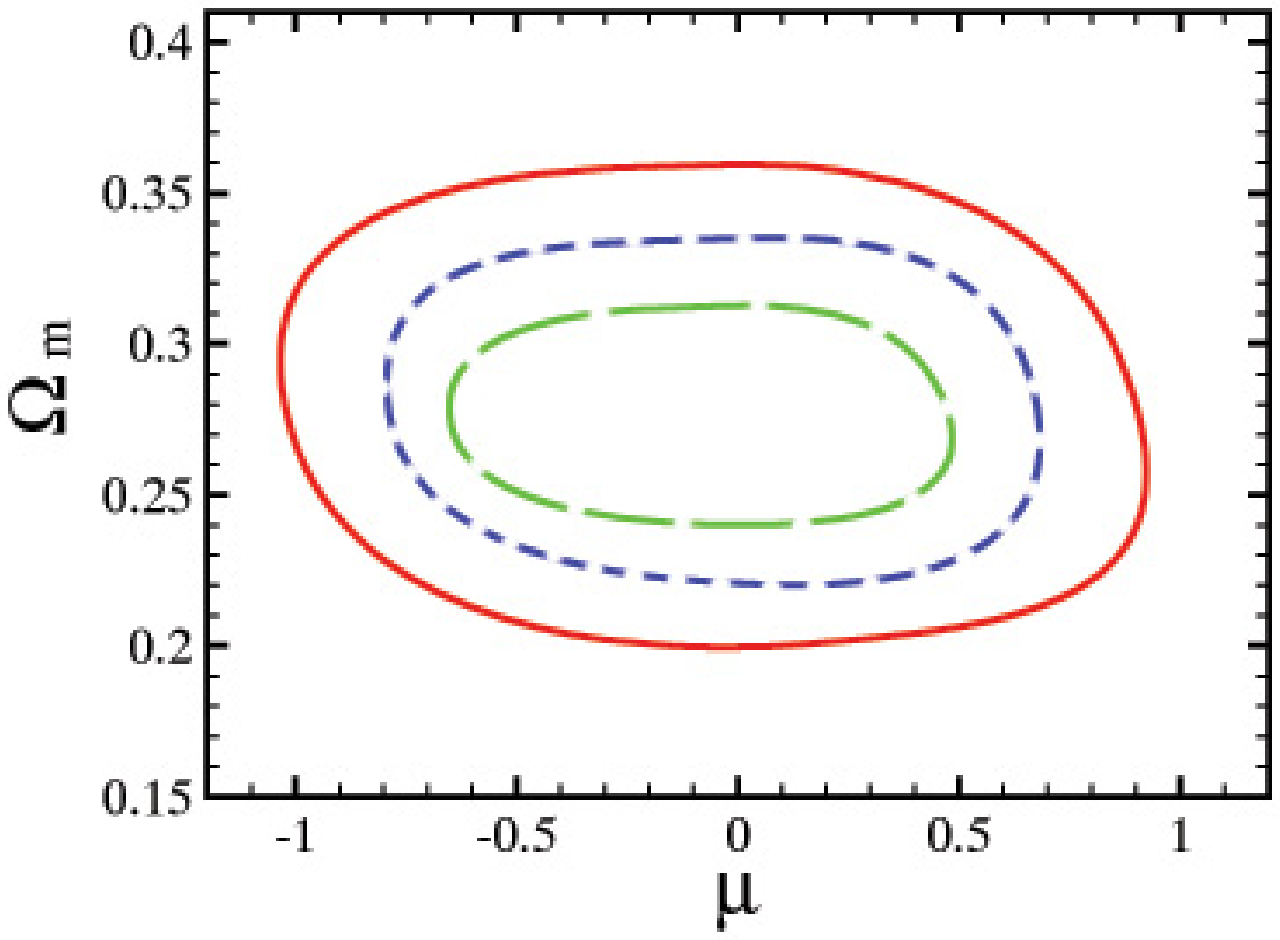}
\caption{Joint confidence intervals of $\mu$ and $\Omega_m$, fitted
with SNIa SNLS$+$CMB$+$SDSS. Solid line, dashed line and long dashed
line correspond to $3\sigma$, $2\sigma$ and $1\sigma$ level of
confidence, respectively.} \label{jlike6}
\end{center}
\end{figure}

In the rest of this subsection we perform a combined analysis of
SNIa, CMB and SDSS to constrain the parameters of the braneworld
model by minimizing the combined $\chi^2 = \chi^2_{\rm
{SNIa}}+\chi^2_{{\rm CMB}}+\chi^2_{{\rm SDSS}}$. The best values of
the model parameters from the fitting with the corresponding error
bars from the likelihood function marginalizing over the Hubble
parameter in the multidimensional parameter space are:
$\Omega_m=0.29_{-0.02}^{+0.03}$, $\Omega_A=-0.71_{-0.03}^{+0.03}$
and $\mu=-0.40^{+0.28}_{-0.26}$ at $1\sigma$
 confidence level with
$\chi^2_{min}/N_{d.o.f}=0.93$ demonstrated
$\Omega_K=+0.00_{-0.04}^{+0.04}$ . The Hubble parameter
corresponding to the minimum value of $\chi^2$ is $H_0=62.72$. Here
we obtain an age of $14.82_{-0.44}^{+0.55}$ Gyr for the Universe
(see section V for more details). Using the SNLS data, the best fit
values of model parameters are: $\Omega_m=0.27_{-0.02}^{+0.02}$
$\Omega_A=-0.74_{-0.02}^{+0.04}$ and $\mu=0.00^{+0.30}_{-0.30}$ at
$1\sigma$ confidence level with $\chi^2_{min}/N_{d.o.f}=0.86$,
states $\Omega_K=-0.01_{-0.03}^{+0.04}$. Age of Universe calculating
with the best fit parameters is $14.05_{-0.45}^{+0.43}$ (see next
section). Tables 2 and 3 give the best fit values for the free
parameters and age of Universe computing with these values. Joint
confidence intervals in free parameter spaces are shown in figures
\ref{jlike1}-\ref{jlike6}.

\begin{table}
\begin{center}
  \begin{tabular}{rccr}
            \hline
            \noalign{\smallskip}

Observation & $\Omega_m$ & $\Omega_A$ & $\mu$
\\ \hline
  &&& \\
 & $0.51^{+0.10}_{-0.30}$&$-0.75_{-1.41}^{+0.32}$&$0.76^{+7.24}_{-1.95}$  \\ 
 SNIa(new Gold)&&&\\
 &  $0.51^{+0.14}_{-0.54}$&$-0.75_{-1.55}^{+0.45}$&$0.76^{+15.24}_{-2.27}$
\\ &&&\\
SNIa(new Gold)+CMB & $0.49^{+0.07}_{-0.08}$&$-0.59^{+0.05}_{-0.06}$& $-1.76 ^{+1.72}_{-1.31}$ \\ 
 &&& \\
&$0.49^{+0.14}_{-0.21}$&$-0.59^{+0.09}_{-0.14}$& $-1.76
^{+4.28}_{-2.32}$
 \\
 &&&\\ 
SNIa(new Gold)+&
$0.29^{+0.03}_{-0.02}$&$-0.71^{+0.03}_{-0.03}$&$-0.40
^{+0.28}_{-0.26}$
 \\
 CMB+SDSS&&&  \\
&$0.29^{+0.05}_{-0.04}$&$-0.71^{+0.06}_{-0.06}$&$-0.40
^{+0.58}_{-0.56}$ \\
 &&&\\  

  & $0.06^{+0.44}_{-0.06}$&$-1.84^{+1.58}_{-0.59}$ & $-1.34 ^{+7.34}_{-0.10}$  \\ 
 SNIa (SNLS)&&&\\
 & $0.06^{+0.68}_{-0.06}$&$-1.84^{+1.73}_{-0.94}$ & $-1.34 ^{+17.34}_{-0.22}$
\\ &&&\\
SNIa(SNLS)+CMB& $0.23^{+0.20}_{-0.23}$&$-0.76^{+0.11}_{-0.16}$ & $-0.26 ^{+1.72}_{-1.00}$ \\
&&& \\
&$0.23^{+0.27}_{-0.23}$&$-0.76^{+0.16}_{-0.17}$ & $-0.26
^{+4.04}_{-1.22}$
 \\
 &&&\\ 
SNIa(SNLS)+& $0.27^{+0.02}_{-0.02}$&$-0.74^{+0.04}_{-0.02}$ & $0.00
^{+0.30}_{-0.30}$
 \\
 CMB+SDSS&&&  \\
&$0.27^{+0.05}_{-0.04}$&$-0.74^{+0.07}_{-0.05}$ & $0.00
^{+0.60}_{-0.60}$ \\
 &&&\\ \hline
            \noalign{\smallskip}
  \end{tabular}
\caption{ The best fit values for the parameters of the model using
SNIa from new Gold sample and SNLS data, SNIa+CMB and SNIa+CMB+SDSS
experiments at one and two $\sigma$ confidence level. }
\end{center}
\label{table2}
\end{table}

\begin{table}
\begin{center}
  \begin{tabular}{rcr}
            \hline
            \noalign{\smallskip}

Observation & $\Omega_K$ & Age (Gyr)
\\\hline
   &&\\
    SNIa(new Gold)&$-0.26^{+0.33}_{-1.44}$&$13.44^{+2.10}_{-7.13}$\\
&&\\

SNIa(new Gold)+CMB & $+0.01^{+0.09}_{-0.10}$&$13.53^{+0.85}_{-1.18}$ \\ 
 && \\
SNIa(new Gold)+&$+0.00^{+0.04}_{-0.04}$&$14.82^{+0.55}_{-0.44}$
 \\
 CMB+SDSS&&  \\
  &&\\
  SNIa (SNLS)&$-0.90^{+1.64}_{-0.59}$&$14.38^{+2.00}_{-1.81}$\\
 && \\
SNIa(SNLS)+CMB& $-0.01^{+0.23}_{-0.28}$&$14.38^{+3.03}_{-14.38}$ \\
&&\\
SNIa(SNLS)+& $-0.01^{+0.04}_{-0.03}$&$14.05^{+0.43}_{-0.45}$ \\
 CMB+SDSS&&  \\
 \hline
            \noalign{\smallskip}
  \end{tabular}
\caption{The best values for the curvature of the brane model with
the corresponding age for the Universe from fitting with SNIa from
new Gold sample and SNLS data, SNIa+CMB and SNIa+CMB+SDSS
experiments at one and two $\sigma$ confidence level. }
\end{center}
\label{table3}
\end{table}

\begin{table}
\begin{center}
  \begin{tabular}{rccr}
            \hline
            \noalign{\smallskip}

     & LBDS &LBDS  & APM\\
    Observation& $53$W$069$&$53$W$091$& $08279+5255$ \\
  & $z=1.43$&$z=1.55$& $z=3.91$  \\
             \hline
           \noalign{\smallskip}
SNIa (new Gold)& $1.00^{+0.10}_{-0.75}$ & $1.06^{+0.10}_{-0.76}$& $0.65^{+0.14}_{-0.41}$ \\
&&&\\
SNIa(new Gold)+CMB &$0.98^{+0.07}_{-0.10}$&$1.04^{+0.07}_{-0.10}$&$0.65^{+0.14}_{-0.07}$ \\
&&&\\
SNIa(new Gold)+CMB & $1.22^{+0.06}_{-0.04}$&$1.31^{+0.06}_{-0.05}$&$0.84^{+0.17}_{-0.04}$ \\
 +SDSS\\
&&&\\ SNIa (SNLS)& $ 2.33^{+1.15}_{-1.13}$ & $2.55^{+1.35}_{-1.31}$& $1.91^{+1.70}_{-1.65}$ \\
&&&\\
SNIa(SNLS)+CMB & $1.23^{+0.31}_{-1.23}$&$1.32^{+0.34}_{-1.32}$&$0.85^{+0.28}_{-0.85}$ \\
&&&\\
SNIa(SNLS)+CMB & $1.16^{+0.04}_{-0.05}$&$1.24^{+0.04}_{-0.05}$&$0.79^{+0.16}_{-0.04}$ \\
 +SDSS \\          \hline
            \noalign{\smallskip}
  \end{tabular}
\caption{ The value of $\tau$ for three high redshift objects, using
the parameters of the model derived from fitting with the
observations. }
\end{center}

\label{table4}
\end{table}

\section{Age of Universe}

\begin{figure}
\begin{center}
\includegraphics[width=\columnwidth]{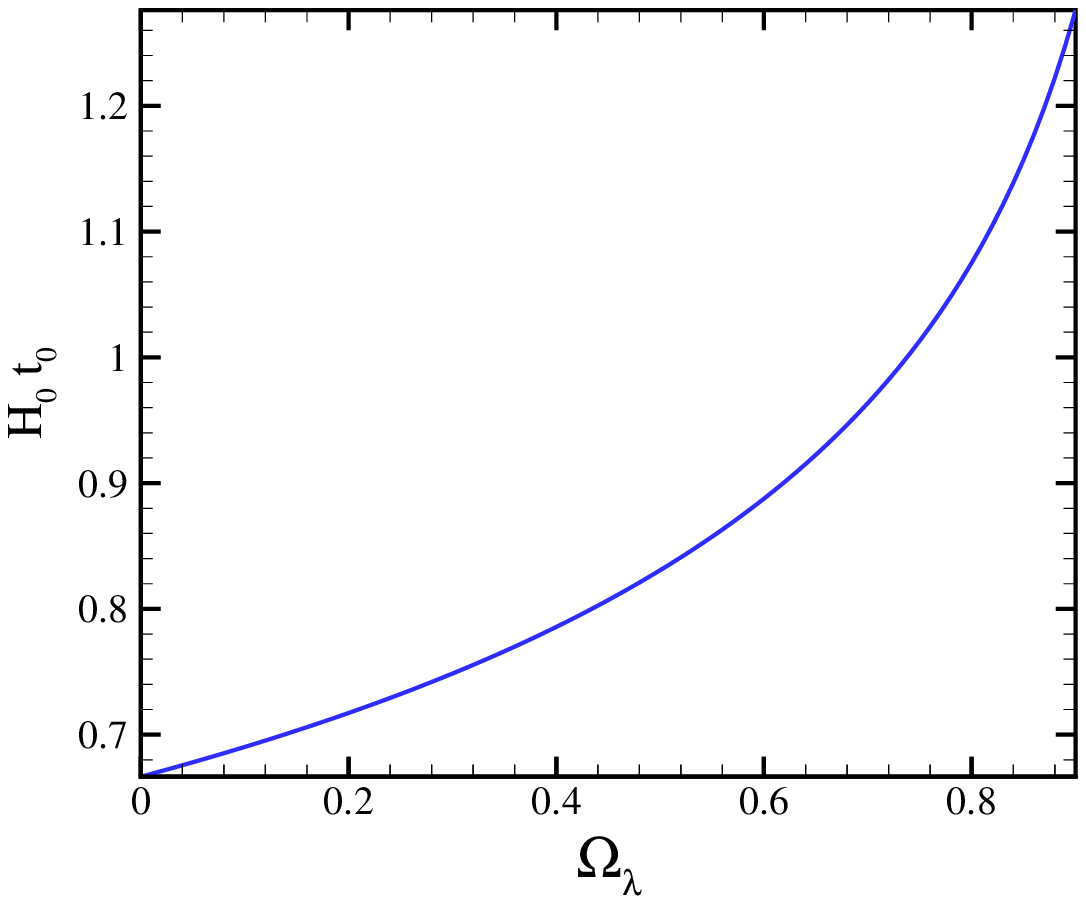}
\includegraphics[width=\columnwidth]{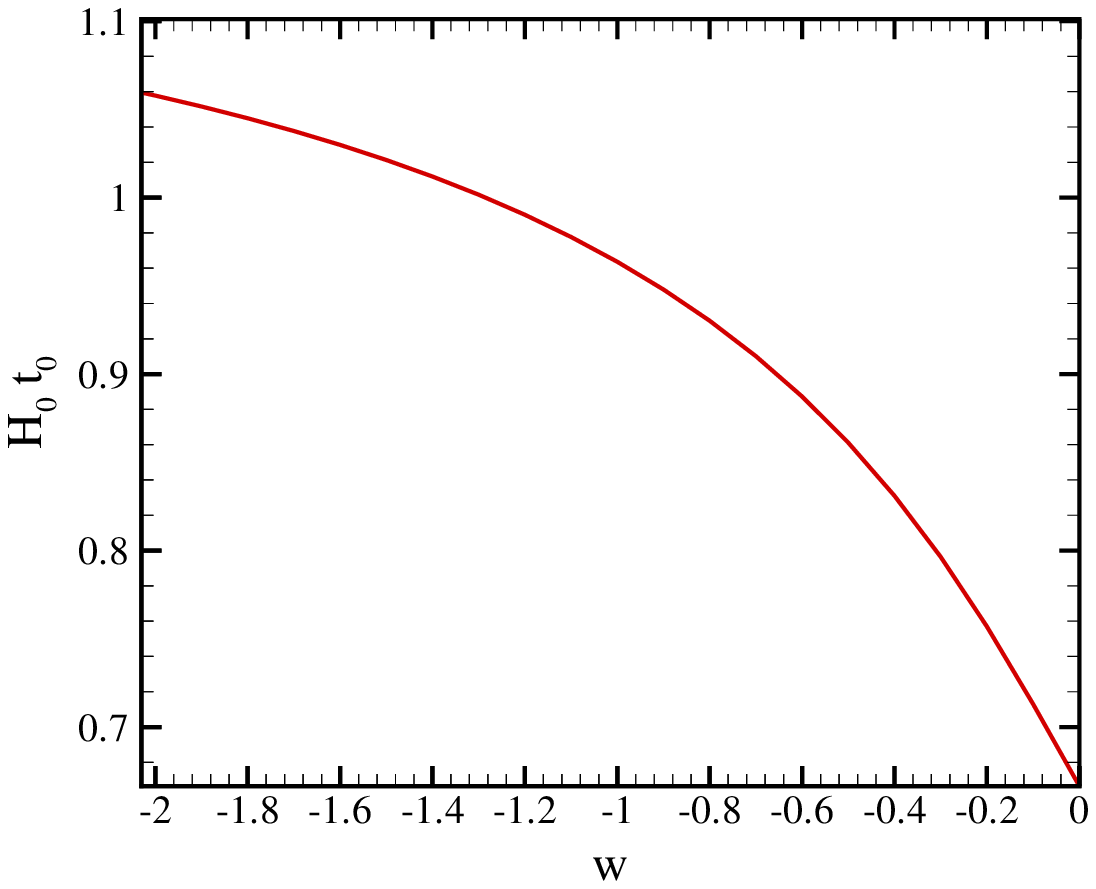}
\caption{ $H_0t_0$ (age of Universe times the Hubble constant at the
present time) versus $\Omega_{\lambda}$ (upper panel) in the
$\Lambda$CDM model for the case $\Omega_K=0.0$ and $w=-1.0$. Lower
panel shows $H_0t_0$ as a function of present equation of state,
$w$, in the $\Lambda$CDM model with the typical values
$\Omega_m=0.30$ and $\Omega_{\lambda}=0.70$.} \label{age2}
\end{center}
\end{figure}

The age of Universe integrated from the big bang up to now in terms
of free parameters of the braneworld model is given by
\begin{eqnarray}\label{age}
&&t_0(\Omega_m,\Omega_A,\mu) = \int_0^{t_0}\,dt \nonumber\\
&&={1 \over H_0\sqrt{|\Omega_K|}}\, {\cal F} \left(
\sqrt{|\Omega_K|}\int_0^\infty {dz'H_0\over (1+z') H(z')} \right).
\end{eqnarray}
Figure \ref{age} shows the dependency of $H_0t_0$ (Hubble parameter
times the age of Universe) on $\Omega_A$ and $\mu$ for a flat
Universe. Obviously increasing $\Omega_A$ and $\mu$ result in a
shorter and longer age for the Universe, respectively. As a matter
of fact, according to the equation (\ref{hub}),  $\Omega_A$ behaves
as inverse role of dark energy in the $\Lambda$CDM scenario and
$\mu$ has the inverse role of $w$ in the $\Lambda$CDM (see Figures
\ref{age} and \ref{age2}).


\begin{figure}
\begin{center}
\includegraphics[width=\columnwidth]{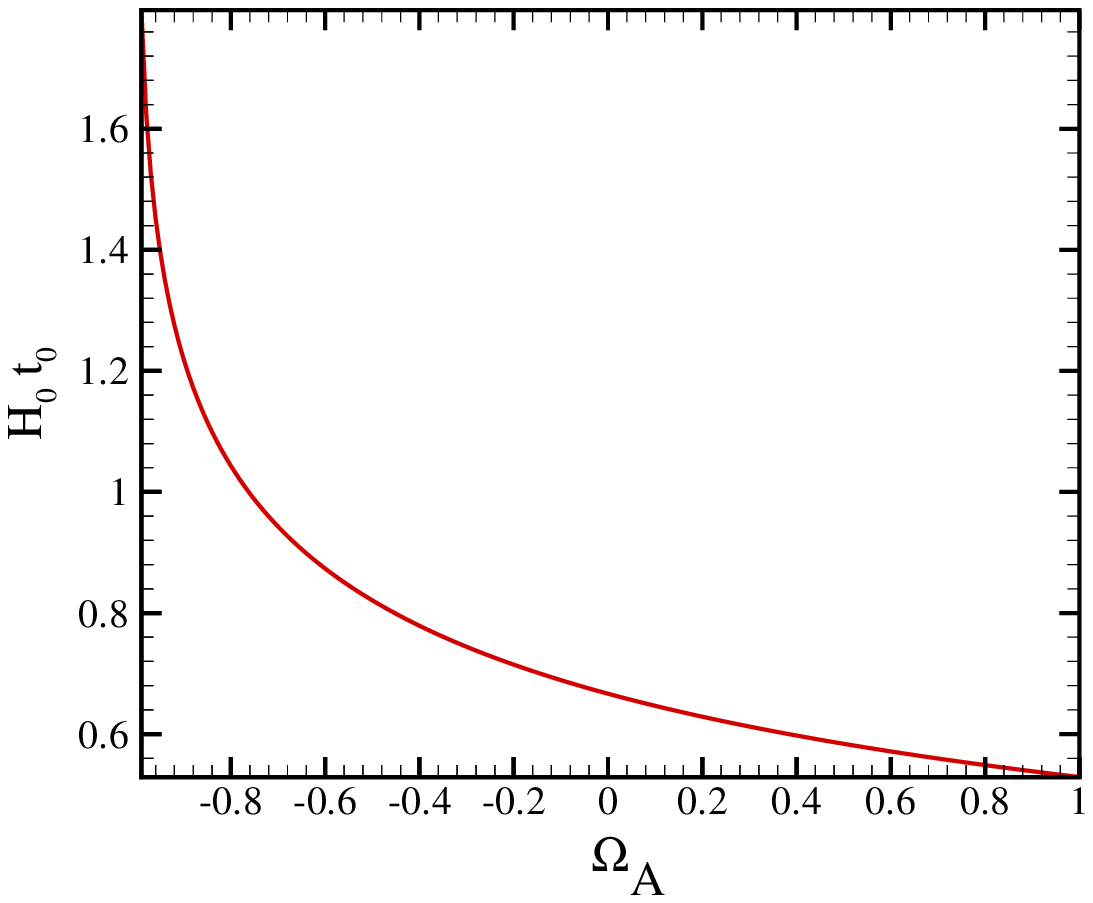}
\includegraphics[width=\columnwidth]{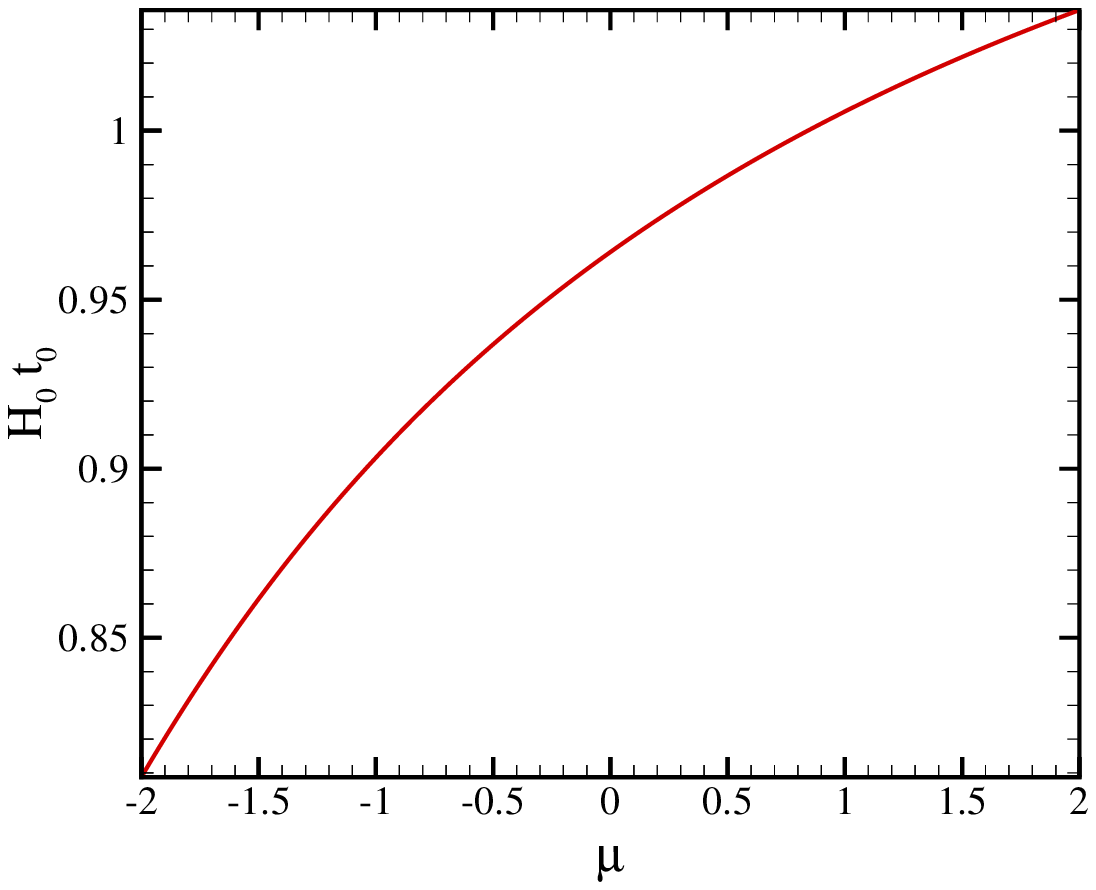}
\caption{ $H_0t_0$ (age of Universe times the Hubble constant at the
present time) as a function of $\Omega_A$ (upper panel) for a flat
Universe and typical value of $\mu=-0.40$. Increasing $\Omega_A$
gives a shorter age for the Universe. Lower panel shows the same
function versus $\mu$ for the case $\Omega_m=0.30$, $\Omega_A=-0.70$
(flat Universe).} \label{age}
\end{center}
\end{figure}

The ``age crisis" is one the main reasons of the acceleration phase
of the Universe. The problem is that the Universe's age in the Cold
Dark Matter (CDM) Universe is less than the age of old stars in it.
Studies on the old stars \citep{carretta00} suggest an age of
$13^{+4}_{-2}$ Gyr for the Universe. Richer et. al. \citep{richer02}
and Hasen et. al. \citep{hansen02} also proposed an age of
$12.7\pm0.7$ Gyr, using the white dwarf cooling sequence method (for
full review of the cosmic age see \citep{spe03}). Table 3 shows that
age of the Universe from the combined analysis of SNIa$+$CMB$+$SDSS
is $14.82_{-0.44}^{+0.55}$ Gyr and $14.05_{-0.45}^{+0.43}$ Gyr for
new Gold sample and SNLS data, respectively, while $\Lambda$CDM
implies $13.7\pm0.2$Gyr \citep{spe03}. These values are in agreement
with the age of old stars \citep{carretta00,krauss1,krauss2}.

To do another consistency test, we compare the age of Universe
derived from this model with the age of old stars and Old High
Redshift Galaxies (OHRG) in various redshifts. Here we consider
three OHRG for comparison with the braneworld model, namely the LBDS
$53$W$091$, a $3.5$-Gyr old radio galaxy at $z=1.55$
\citep{dunlop96,spin}, the LBDS $53$W$069$ a $4.0$-Gyr old radio
galaxy at $z=1.43$ \citep{dunlop99} and a quasar, APM $08279+5255$
at $z=3.91$ with an age of $t=2.1_{-0.1}^{+0.9}$Gyr
\citep{hasinger02,komo02}. The later has once again led to the "age
crisis". An interesting point about this quasar is that it cannot be
accommodated in the $\Lambda$CDM model \citep{jan06}. In order to
quantify the age-consistency test we introduce the expression $\tau$
as:
\begin{equation}
 \tau=\frac{t(z;\Omega_m,\Omega_A,\mu)}{t_{{\rm obs}}} = \frac{t(z;\Omega_m,\Omega_A,\mu)H_0}{t_{{\rm
 obs}}H_0},
\end{equation}
where $t(z)$ is the age of Universe, obtained from the equation
(\ref{age}) and $t_{obs}$ is an estimation for the age of old
cosmological object. In order to have a compatible age for the
Universe we should have $\tau>1$. Table 4 reports the value of
$\tau$ for three mentioned OHRG with various observations. We see
that the parameters of braneworld model from the combined
observations provide a compatible age for the Universe, compared to
the age of old objects, also in addition, SNLS data result in a
shorter age for the Universe. Once again for the braneworld model,
APM $08279+5255$ at $z=3.91$ has a longer age than the Universe but
gives better result than most cosmological models investigated
before \citep{sa1,sa2,jan06}.

\section{conclusions and discussions}

The impressive amount of data indicating a spatially flat Universe
in accelerated expansion has posed the problem of dark energy and
stimulated the search for cosmological models which are able to
explain such unexpected behavior. Many rival theories have been
proposed to solve the puzzle of the nature of dark energy ranging
from a rolling scalar field to a unified picture where a single
exotic fluid accounts for the whole dark sector (dark matter and
dark energy). Moreover, modifications of the gravity Lagrangian have
also been advocated. Although deeply different in their underlying
physics, all these scenarios share the common feature of well
reproducing the available astrophysical data. On the other hand,
alternative cosmology from the braneworld models provide a possible
mechanism for the present acceleration of the Universe congruously
suggested by various cosmological observations.

In braneworld scenarios, due to the usual energy conservation law on
the brane, we do not have energy flow from the brane onto the bulk
or vice versa. There are numerous efforts to constrain the
braneworld models but in all of them, there is no energy exchange
between the brane and bulk. Theoretically, there are no fundamental
reasons to forbid the energy exchange between the brane and bulk in
a brane scenario. One can get this profile by relaxing the
conservation law on the brane. This energy exchange can alter the
profile of the cosmic expansion and leads to a behavior that would
resemble the dark energy. In this paper we focused our attention on
the RS II braneworld model with energy exchange between the brane
and bulk. We got the modified Friedmann equation (\ref{hub}) on the
brane which can explain the cosmological behavior and describe a
physically origin for the dark energy which is in good agreement
with observations. We explored the consistency of this scenario with
the implication of up-to-date luminosity of supernova type Ia
observed by two independent groups, new Gold sample and SNLS data
set, acoustic peak in the cosmic microwave background anisotropy
power spectrum and baryon acoustic oscillation measured by Sloan
Digital Sky Survey. The effect of model free parameters on the
matter power spectrum and the exploration of matter and dark energy
interaction will investigate in our forthcoming paper.

The best parameters obtained from the fitting with the new Gold
sample data combined with CMB and SDSS observations are:
$\Omega_m=0.29_{-0.02}^{+0.03}$, $\Omega_A=-0.71_{-0.03}^{+0.03}$
and $\mu=-0.40^{+0.28}_{-0.26}$ at $1\sigma$ confidence level with
$\chi^2_{min}/N_{d.o.f}=0.93$ expressing spatially flat Universe
with $\Omega_K=+0.00_{-0.04}^{+0.04}$. SNLS SNIa$+$CMB$+$SDSS give:
$\Omega_m=0.27_{-0.02}^{+0.02}$ $\Omega_A=-0.74_{-0.02}^{+0.04}$ and
$\mu=0.00^{+0.30}_{-0.30}$ at $1\sigma$ confidence level with
$\chi^2_{min}/N_{d.o.f}=0.86$, asserting
$\Omega_K=-0.01_{-0.03}^{+0.04}$. The well known $\Lambda$CDM model
implying $-0.06\le\Omega_K\le+0.02$ \citep{spe03} and  some other
interesting models such as Dvali-Gabadadze-Porrati (DGP) which
indicates $\Omega_K=0.01^{+0.09}_{-0.09}$ and
$\Omega_K=0.01^{+0.04}_{-0.04}$ using Gold sample and SNLS data,
respectively \citep{0603632v2,sadgp,sima}.

We also performed the age test, comparing the age of old stars and
old high redshift galaxies with the age derived from this model.
From the best fit parameters of the model using new Gold sample and
SNLS SNIa,  we obtained an age of $14.82_{-0.44}^{+0.55}$ Gyr and
$14.05_{-0.45}^{+0.43}$ Gyr, for the Universe, respectively. These
results are in agreement with the age of the old stars. The age of
Universe in this model is larger than what given in the other models
\citep{spe03,sa1,sa2,sadgp}.

To check the age crisis in this model we chose two high redshift
radio galaxies at $z=1.55$ and $z=1.43$ with a quasar at $z=3.91$.
Two first objects were consistent with the age of Universe, i.e.,
they were younger than the Universe while the third one was not but
our model gave the better result than $\Lambda$CDM and a class of
Quintessence model \citep{sa1,sa2}.

Finally, it must point out that the energy exchange term $\Omega_A$
plays a crucial role in our work. In other words, in the RS II model
without energy exchange where we have $\Omega_A=0$, we can not get
late time acceleration expansion profile for our Universe! So we
conclude that the usual RS II model should ruled out from present
observational data.

\section*{Acknowledgements}
We would like to thank the anonymous referee for his/her useful
comments. Also authors is grateful to Mr Mohammadi Najafabadi for
reading the manuscript and useful comments. A. Sheykhi thanks Bin
Wang for his valuable suggestions and helpful discussions. This work
was partially supported by Shahid Bahonar University of Kerman.


\bsp \label{lastpage}

\begin{thebibliography}{99}
\bibitem[\protect\citeauthoryear{Aaquist}{1993}]{aaquist93} Aaquist, O.~B.\  1993, A\& A, 267, 260
\bibitem[\protect\citeauthoryear{Aaquist \& Kwok}{1989}]{aaquist89} Aaquist, O.~B. \& Kwok, S. 1989, A\& A, 222,
227.

\bibitem[\protect\citeauthoryear{M. Ahmed et. al.}{2004}]{ahmad}  Ahmed, M.,
Dodelson, S.,  Greene,  P. B.  and  Sorkin, R., Phys. Rev. D 69
103523(2004).
\bibitem[\protect\citeauthoryear{C. Alcock el. al.}{1979}]{alc79} Alcock, C. and  Paczynski, B., Nature  281, 358 (1979).
\bibitem[\protect\citeauthoryear{ M. Amarzguioui et. al.}{2005}]{Amarzguioui}  Amarzguioui, M.,  Elgaroy, O.,  Mota, D. F.,
Multamaki, T., 2006, A\&A, 454, 707.
\bibitem[\protect\citeauthoryear{L. Amendola }{2003a}]{amen3} Amendola, L., Mon. Not. R. Astron. Soc. 342, 221 (2003a).
\bibitem[\protect\citeauthoryear{L. Amendola}{2000}]{LAmendola2000}Amendola, L., Phys. Rev. D {\bf 62}, 043511 (2000).
\bibitem[\protect\citeauthoryear{L. Amendola et. al.}{2001}]{LAmendola2001}Amendola, L. and  Tocchini-Valentini,
D., Phys. Rev. D {\bf 64}, 043509 (2001).
\bibitem[\protect\citeauthoryear{L. Amendola}{2002}]{LAmendola2002}Amendola, L. and  Tocchini-Valentini,
D., Phys. Rev. D {\bf 66}, 043528 (2002).
\bibitem[\protect\citeauthoryear{L. Amendola et. al.}{2003b}]{LAmendola2003} Amendola, L.,  Quercellini, C.,   Tocchini-Valentini, D. and
Pasqui, A., Astrophys. J. {\bf 583}, L53 (2003b).


\bibitem[\protect\citeauthoryear{P. S. Apostolopoulos et. al.} {2005}]{apos1}  Apostolopoulos, P. S.,
Tetradis,  N.,  Phys. Rev. D 71 (2005) 043506.

\bibitem[\protect\citeauthoryear{P. S. Apostolopoulos et. al.}{2006}]{apos2}  Apostolopoulos, P. S. and  Tetradis, N., Phys. Lett. B
633 409 (2006).
\bibitem[\protect\citeauthoryear{S. Arbabi-Bidgoli el. al.}{2006}]{arb05} Arbabi-Bidgoli, S., Movahed, M. S. and  Rahvar, S., International
Journal of Modern Physics D Vol.  15, No. 9 (2006) 1455–1472.

\bibitem[\protect\citeauthoryear{N. Arkani-Hamed et. al.}{2002}]{arkani}  Arkani-Hamed, N.,
Dimopoulos,  S.,  Dvali,  G.  and  Gabadadze, G.,  hep-th/0209227.

\bibitem[\protect\citeauthoryear{C. Armendariz-Picon el. al.}{2000}]{arm00} Armendariz-Picon, C.,  Mukhanov, V.  and  Steinhardt, P. J., Phys. Rev.
Lett.  85, 4438 (2000).
\bibitem[\protect\citeauthoryear{P. Astier el. al.}{2005}]{astier05}  Astier, P., et al., 2006, A\&A. 447, 31.


\bibitem[\protect\citeauthoryear{B. J. Barris el. al.}{2004}]{bar04}  Barris, B. J. et al., Astrophys. J.  602, 571 (2004).
\bibitem[\protect\citeauthoryear{J. S. Bagla el. al.}{2003}]{pad03} Bagla, J. S.,  Jassal, H. K.  and  Padmanabhan, T., Phys. Rev. D  67,
063504 (2003).


\bibitem[\protect\citeauthoryear{Shant Baghram el. al.}{2007}]{rahvar07}  Baghram, S.,  Farhang,  M. and  Rahvar, S.,  Phys. Rev. D  75, 044024 (2007).
\bibitem[\protect\citeauthoryear{C. L. Bennett el. al.}{2003}]{bennett}  Bennett C. L., et al., Astrophys. J. Suppl. Ser. 148, 1 (2003).
\bibitem[\protect\citeauthoryear{M. C. Bento el. al.}{2002}]{be02}  Bento, M. C.,   Bertolami, O. and  Sen, A. A., Phys. Rev. D 66, 043507 (2002).
\bibitem[\protect\citeauthoryear{ P.~Binetruy el. al. }{2000}]{Bin} Binetruy, P.,  C.~Deffayet and D.~Langlois, Nucl. Phys. B
565 (2000) 269.
\bibitem[\protect\citeauthoryear{C. Blake el. al.}{2003}]{blak03}  Blake, C. and K. Glazebrook, Astrophys. J.  594, 665 (2003)
\bibitem[\protect\citeauthoryear{Balick \& Frank}{2002}]{balick02} Balick, B., \& Frank, A.\ 2002, ARA\&A, 40,
439.

\bibitem[\protect\citeauthoryear{Blackman et al.}{2001}]{black01} Blackman, E.~G., Frank, A., Markiel, J.~A., Thomas, J.~H., \& Van
Horn, H.~M.\ 2001, Nature, 409, 485.
\bibitem[\protect\citeauthoryear{C. Bogdanos el. al.}{2007}]{Bog1}   Bogdanos, C.  and
Tamvakis, K., Phys.Lett. B646 (2007) 39-46.
\bibitem[\protect\citeauthoryear{ J. R. Bond el. al.}{1997}]{bond97}  Bond, J. R.,  Efstathiou, G. and  Tegmark, M.,  Mon. Not. R. Astron.
Soc. 291, L33 (1997).
\bibitem[\protect\citeauthoryear{C. Bogdanos et. al.}{2006}]{Bog2}  Bogdanos, C., Dimitridis, A.,  Tamvakis, K.,  hep-th/0611094.
\bibitem[\protect\citeauthoryear{R.G. Cai el. al.}{2006}]{Cai}  Cai, R.G.,   Gong, Y. and  Wang, B., JCAP 0603, (2006) 006
.
\bibitem[\protect\citeauthoryear{R. R. Caldwell el. al.}{2003}]{cal03}  Caldwell, R. R.,  Kamionkowski, M. and  Weinberg, N. N., Phys. Rev. Lett.
 91, 071301 (2003).
\bibitem[\protect\citeauthoryear{ R. R. Caldwell et. al.}{1998}]{Caldw}  Caldwell, R. R.,   Dave, R. and  Steinhardt, P. J., Phys.
Rev. Lett. 80, 1582 (1998).

\bibitem[\protect\citeauthoryear{R. R. Caldwell el. al.}{2004}]{dor04}  Caldwell, R. R. and Doran, M.,  Phys. Rev. D  69, 103517 (2004).
\bibitem[\protect\citeauthoryear{S. Capozziello et. al.}{2004}]{Capo04} Capozziello, S.,  V. F. Cardone, M. Funaro, S. Andreon, Phys. Rev. D 70, 123501 (2004).

\bibitem[\protect\citeauthoryear{R. R. Caldwell}{2002}]{cal02}  Caldwell, R. R.,  Phys. Lett. B  545, 23 (2002).
 \bibitem[\protect\citeauthoryear{S. M. Carroll}{2001}]{carr} Carroll,  S. M., Living Rev. Relativity 4, 1 (2001).
 \bibitem[\protect\citeauthoryear{E. Carretta el. al.}{2000}]{carretta00} Carretta E., Gratton R., Clementini G. and Fusi Pecci F. 2000, ApJ 533,
 215.
\bibitem[\protect\citeauthoryear{B. Chaboyer et. al. }{2002}]{krauss2} Chaboyer, B. and L. M. Krauss, Astrophys. J. Lett. 567, L45 (2002).
\bibitem[\protect\citeauthoryear{L. P. Chimento and D. Pavon}{2006}]{LPChimento2006} Chimento, L. P. and D. Pavon, Phys. Rev. D {\bf 73}, 063511 (2006).
\bibitem[\protect\citeauthoryear{T. Clifton el. al.}{2005}]{clif}  Clifton, T.,  Barrow, J. D.,  Phys. Rev. D 72, 103005 (2005).
\bibitem[\protect\citeauthoryear{D. Comelli et. al.}{2003}]{come3} Comelli, D.,  Pietroni,M.,  and Riotto, A., Phys. Lett. B 571, 115 (2003).
\bibitem[\protect\citeauthoryear{E. J. Copeland el. al.}{2006}]{cop06} Copeland, E. J.,   Sami, M. and  Tsujikawa, S., Int.J.Mod.Phys. D15 (2006) 1753-1936.

\bibitem[\protect\citeauthoryear{ M. P. Dabrowski el. al.}{2003}]{da03}  Dabrowski, M. P., Stachowiak, T.  and  Szydlowski, M., Phys. Rev. D 68, 103519 (2003).
\bibitem[\protect\citeauthoryear{ M. P. Dabrowski el. al.}{2004}]{da04} Dabrowski, M. P., Godlowski, W. and Szydlowski, M.,
 Gen. Rel. Grav. 36, (2004) 767.

\bibitem[\protect\citeauthoryear{C. Deffayet et. al.}{2002}]{Deff}  Deffayet, C.,   Dvali, G. and  Gabadadze, G., Phys.
Rev. D 65, 044023 (2002).
\bibitem[\protect\citeauthoryear{S. Dodelson}{2003}]{dodelson} Dodelson, S., "Modern Cosmology", Academic Perss,
2003.


\bibitem[\protect\citeauthoryear{M. Doran el. al.}{2001}]{dor01} Doran, M.,  Lilley, M.,  Schwindt, J.  and Wetterich, C., Astrophys. J. 559,
501 (2001).
\bibitem[\protect\citeauthoryear{ Michael Doran el. al.}{2001}]{doran01} Doran, M. and  Lilley, M., Mon.Not.Roy.Astron.Soc. 330 (2002) 965-970.
\bibitem[\protect\citeauthoryear{J. Dunlop el. al.}{1996}]{dunlop96} Dunlop J., Peacock J., Spinrad H., Dey A., Jimenez R., Stern D. and  Windhorst R., Nature (London)
381, 581 (1996).


\bibitem[\protect\citeauthoryear{J. Dunlop}{1999}]{dunlop99}  Dunlop, J., in {\it The Most Distant Radio Galaxies}, edited by H. J.
A. Rottgering, P. Best, and M. D. Lehnert (Kluwer, Dordrecht,
1999), p. 71.


\bibitem[\protect\citeauthoryear{G. Dvali et. al.}{2003}]{Dvali02}  Dvali, G.  and  Turner, M. S., Fermilab pub.
03040-A (2003).


\bibitem[\protect\citeauthoryear{G.~R.~Dvali et. al.}{2000}]{Dvali} Dvali, G.~R., Gabadadze, G., and Porrati, M., Phys.  Lett.  B  484, 112 (2000).


\bibitem[\protect\citeauthoryear{D. J. Eisenstein el. al.}{ 2005}]{eisenstein05}  Eisenstein, D. J., et al., Astrophys.J. 633 (2005) 560-574.
\bibitem[\protect\citeauthoryear{ D. J. Eisenstein et. al.}{1998}] {Eisen98}  Eisenstein, D. J. and W. Hu, Astrophys. J. 496 (1998) 605.

\bibitem[\protect\citeauthoryear{D. Eisenstein et. al. }{2004}]{eisenstein04}  Eisenstein, D. and  White, M. J., Phys. Rev. D  70 103523 (2004).

\bibitem[\protect\citeauthoryear{V. R. Eke et. al.}{2001}]{vink00}  Eke, V. R. et. al., Astrophys.J. 554 (2001) 114-125.


\bibitem[\protect\citeauthoryear{ W. Fischler el. al.}{2001}]{Fis} Fischler W., Kashani-Poor A.,  McNess P. and  Paban S., J. High Energy Phys. 07 (2001) 003.
\bibitem[\protect\citeauthoryear{Franca et. al.}{2004}]{fran4} Franca, U., and Rosenfeld, R., Phys. Rev. D 69, 063517 (2004).
\bibitem[\protect\citeauthoryear{J. A. Frieman et. al.}{1995}]{Friem}  Frieman, J. A.,  Hill,  C. T.,   Stebbins, A., and  Waga, I., Phys. Rev. Lett.
75, 2077 (1995).
\bibitem[\protect\citeauthoryear{K. Freese et. al.}{2002}]{freese}  Freese, K.  and  Lewis, M., Phys. Lett. B 540, 1 (2002).


\bibitem[\protect\citeauthoryear{A. Friaca el. al.}{2005}]{Fri} Friaca, A.,  Alcaniz, J. S.,  Lima,  J. A. S., Mon. Not. Roy. Astron. Soc. 362 (2005) 1295.
\bibitem[\protect\citeauthoryear{W. L. Freedman el. al.}{2001}]{hst} Freedman, W. L. et al., Astrophys. J. Lett. 553, 47 (2001).


\bibitem[\protect\citeauthoryear{S. Ghassemi el. al.}{2006}]{ghkhm} Ghassemi, S., S. Khakshournia, R. Mansouri, J. High Energy Phys. 08 (2006) 019,
(gr-qc/0605094).

\bibitem[\protect\citeauthoryear{Zong-Kuan Guo el. al.}{2006}]{0603632v2} Guo,Z. K.  el. al., arXiv:astro-ph/0603632.
\bibitem[\protect\citeauthoryear{Zong-Kuan Guo el. al.}{2006}]{zong}  Guo Z. K., Zhu Z. H., Alcaniz J.S., Zhang Y. Z. Astrophys.J. 646
(2006) 1.

\bibitem[\protect\citeauthoryear{Wayne Hu el. al.}{1997}]{hu96}  Hu, W.,  Naoshi Sugiyama  and  Joseph Silk, Nature 386 (1997) 37.

\bibitem[\protect\citeauthoryear{W. Hu el. al.}{1995}]{Hu95} Hu, W.  and N. Sugiyama, Astrophys. J.  444, 489 (1995).
\bibitem[\protect\citeauthoryear{ Hu, W. et. al.}{2001}]{Hu} Hu, W., Fukugita, M., Zaldarriaga, M. and Tegmark, M. Astrophysical Journal 549 (2001) 669.
\bibitem[\protect\citeauthoryear{C. Heymans et. al.}{2005}] {Heyman}  Heymans, C.,  et. al. MNRAS 361 (2005) 160.
\bibitem[\protect\citeauthoryear{ W. Hu et. al.}{1996}]{Hu96} Hu, W.  and N. Sugiyama, Astrophys. J. 471 (1996) 542.
\bibitem[\protect\citeauthoryear{B. M. S. Hansen el. al.}{2002}]{hansen02}  Hansen, B. M. S. et al., Astrophys. J. 574, L155 (2002).

\bibitem[\protect\citeauthoryear{G. Hasinger el. al.}{2002}]{hasinger02}  Hasinger, G.,  N. Schartel and S. Komossa, Astrophys. J. Lett.  573,
L77 (2002).
\bibitem[\protect\citeauthoryear{D. Jain. el. al.}{2005}]{jan06}  Jain D., Dev A., Phys.Lett. B633 (2006) 436-440.


 \bibitem[\protect\citeauthoryear{H. K. Jassal el. al.}{2006}]{pad06}  Jassal H. K.,  Bagla J. S. and  Padmanabhan T., astro-ph/0601389.
\bibitem[\protect\citeauthoryear{D. Jeong and E. Komatsu}{2006}]{donc06}Jeong, D. and Komatsu, E., Astrophys.J. 651 (2006) 619.



\bibitem[\protect\citeauthoryear{T. koivisto}{2006}]{tomikoivisito06}, Koivisto, T., Phys.Rev. D73 (2006) 083517.

\bibitem[\protect\citeauthoryear{L. M. Krauss et. al.}{2001}]{krauss1} Krauss, L. M. and B. Chaboyer, astro-ph/0111597.
\bibitem[\protect\citeauthoryear{A. Kamenshchik el. al.}{2001}]{kam01}  Kamenshchik, A.,   Moschella, U. and  Pasquier, V., Phys. Lett. B  511,
265 (2001).
\bibitem[\protect\citeauthoryear{E. Kiritsis el. al.}{2002}]{Kirit2}  Kiritsis, E.,  Tetradis, N. and  Tomaras, T. N., JHEP 0203
(2002) 019.

\bibitem[\protect\citeauthoryear{E. Kiritsis} {2005}]{Kirit3}  Kiritsis, E.,  JCAP 0510 014
(2005).
\bibitem[\protect\citeauthoryear{E. Kiritsis el. al.} {2003}] {Kirit1}
Kiritsis, E.,   Kofinas, G.,  Tetradis, N.,  Tomaras,  T. N. and
 Zarikas, V., JHEP 0302 (2003) 035.


\bibitem[\protect\citeauthoryear{S. Komossa et. al.}{2002}]{komo02}  Komossa, S. and G. Hasinger, to appear in the proc. of the workshop "XEUS - studying the evolution of the universe", G. Hasinger et al. (eds), MPE Report, in
press, astro-ph/0207321.
\bibitem[\protect\citeauthoryear{ L. Knox el. al.}{2001}]{Knox} Knox, L., Christensen, N. and Skordis, C. Astrophysical
Journal 563 (2001) L95.
\bibitem[\protect\citeauthoryear{C. L. Kuo et. al.}{2004}]{Kuo}  Kuo, C. L., et al. (ACBAR Collaboration), Astrophys. J. 600, 32 (2004).

\bibitem[\protect\citeauthoryear{ G. Kofinas el. al.}{2005}]{Kof}  Kofinas, G., Panotopoulos,  G. and  Tomaras, T.N., JHEP 0601 (2006) 107.

\bibitem[\protect\citeauthoryear{S. Lee et. al.}{2006}]{seok06} Lee S., Liu Guo-Chin and  Ng
Kin-Wang, Phys.Rev. D73 (2006) 083516.
\bibitem[\protect\citeauthoryear{ J. A. S. Lima }{2004}]{lim04}  Lima,  J. A. S.,  Braz. J. Phys. 34, 194 (2004).

\bibitem[\protect\citeauthoryear{ A. R. Liddle}{1998}]{Liddle}  Liddle, A. R.  and  Scherrer, R. J., Phys.
Rev. D 59, 023509 (1998).

\bibitem[\protect\citeauthoryear{Li, M.}{2004}]{li04} Li, M., Phys.Lett. B, 603, 1 (2004).
\bibitem[\protect\citeauthoryear{E. V. Linder}{2005}]{linder1} Linder, E. V., arXiv:astro-ph/0507308.

\bibitem[\protect\citeauthoryear{E. V. Linder}{2003}]{linder2}  Linder, E. V., Phys. Rev. D  68, 083504 (2003).
\bibitem[\protect\citeauthoryear{A. Loeb and M. Zaldarriaga}{2005}]{abraham05} Loeb, A. and  Zaldarriaga,
M., Phys.Rev. D71 (2005) 103520.
\bibitem[\protect\citeauthoryear{C. P. Ma et. al.}{1999}]{chungpeima99} Ma, C. P.,  Caldwell, R. R.,  Bode, P. and  Wang,
L., Astrophys.J. 521 (1999) L1-L4.
\bibitem[\protect\citeauthoryear{Z. Ma}{2006}]{zhaominma07}  Ma, Z., The Astrophysical Journal, 2007, Volume 665, Issue 2, pp. 887-898, arXiv:astro-ph/0610213.

\bibitem[\protect\citeauthoryear{Masciadri et al.}{2002}]{masciadri02} Masciadri, E., de Gouveia Dal Pino, E.~M., Raga, A.~C., \& Noriega-Crespo, A.\ 2002,
ApJ, 580, 950
\bibitem[\protect\citeauthoryear{A. Melchiorri el. al.}{2003}]{bbn} Melchiorri, A., L. Mersini, C.L. $\ddot{O}$dman and M.
Trodden, Phys.Rev. D68 (2003) 043509.
\bibitem[\protect\citeauthoryear{Mellema}{1995}]{mellema95} Mellema, G.\ 1995,  MNRAS, 277,
173.

\bibitem[\protect\citeauthoryear{Miranda et al.}{1998}]{lfm98} Miranda, L.~F., Fernandez Matilde, Alcala Juan M.,
 Guerrero, Martin A., Anglada Guillem, Gomez Yolanda, Torrelles, José M. and Aaquist Orla B. 2000, MNRAS, 311,
748.

\bibitem[\protect\citeauthoryear{Miranda et al.}{2000}]{lfm00} Miranda, L.~F., Torrelles, J.~M., Guerrero, M.~A., Aaquist, O.~B., \& Eiroa, C.\
1998, MNRAS, 298, 243.
\bibitem[\protect\citeauthoryear{ L.~F. Miranda et al.}{2001}]{sa1} Miranda, L.~F., G{\'o}mez, Y., Anglada, G., \& Torrelles, J.~M.\ 2001, Nature, 414,
284.
\bibitem[\protect\citeauthoryear{Miranda et al.}{2001}]{lfm01} Miranda, L.~F., G{\'o}mez, Y., Anglada, G., \& Torrelles, J.~M.\ 2001, Nature, 414,
284.

\bibitem[\protect\citeauthoryear{Morris}{1987}]{morris87} Morris, M.\ 1987, PASP, 99,
1115.
\bibitem[\protect\citeauthoryear{M. Sadegh Movahed el. al.}{2007}]{sar07} Movahed, M. S., Baghram, S. and Rahvar, S., Phys. Rev. D  76, 044008 (2007).

\bibitem[\protect\citeauthoryear{Movahed et. al.}{2007}]{sadgp} Movahed, M. S., Farhang M. and Rahvar S., arXiv:astro-ph/0701339.

\bibitem[\protect\citeauthoryear{Movahed and Gassemi}{2007}]{sima}  Movahed, M. S., Ghassemi, S., Phys. Rev.  D {\bf 76} 084037 (2007).

\bibitem[\protect\citeauthoryear{S. Nesseris el. al.}{2004}]{Nesseris04}  Nesseris, S. and  Perivolaropoulos, L., Phys. Rev.  D 70,
043531 (2004).
\bibitem[\protect\citeauthoryear{S. Nesseris el. al.}{2007}]{ness06}  Nesseris, S. and  Perivolaropoulos,
L., Journal of Cosmology and Astroparticle Physics, 0701 (2007) 018.
\bibitem[\protect\citeauthoryear{S. Nojiri et. al. }{2003a}]{noji1}  Nojiri, S. and  Odintsov, S. D.,  Phys.
Rev. D 68, 123512 (2003a).

\bibitem[\protect\citeauthoryear{S. Nojiri et. al.}{2003b}] {noji2}  Nojiri, S., and  Odintsov, S. D., Phys. Lett. B 562, 147 (2003b).

\bibitem[\protect\citeauthoryear{S. Nobili et. al.}{2005}]{Nob05} Nobili, S., et. al., Astronomy and Astrophysics 437 (2005) 789.


\bibitem[\protect\citeauthoryear{Osterbrock}{1989}]{osterbrock89} Osterbrock, D.~E.\ 1989, Research supported by the University of California, John Simon
Guggenheim Memorial Foundation, University of Minnesota, et
al.~Mill Valley, CA, University Science Books, pp. 86-96.
\bibitem[\protect\citeauthoryear{ C. J. Odman et. al. }{2003}]{Odman}  Odman, C. J.,   Melchiorri, A.,  Hobson, M. P. and  Lasenby, A. N., Phys. Rev. D 67,
083511 (2003).

\bibitem[\protect\citeauthoryear{G. Olivares et. al.}{2005}]{german05}  Olivares,
G.,Atrio-Barandela F. and Pavon D.,Phys.Rev. D71 (2005) 063523.
\bibitem[\protect\citeauthoryear{G. Olivares et. al.}{2006}]{german06} Olivares, G., Atrio-Barandela F. and Pavon D., Phys.Rev. D74 (2006) 043521.


\bibitem[\protect\citeauthoryear{T. Padmanabhan}{2003}]{padm}  Padmanabhan, T., Phys. Rep.  380, 235 (2003).
\bibitem[\protect\citeauthoryear{L. Page el. al.}{2003}]{pag03}  Page, L., et al., Astrophys. Supp. J.  148, 233 (2003).
\bibitem[\protect\citeauthoryear{T. J. Pearson el. al.}{2003}]{pearson03}  Pearson, T. J., et al. (CBI Collaboration), Astrophys. J. 591, 556
(2003).
\bibitem[\protect\citeauthoryear{W. J. Percival et. al. }{2002}]{Perci}  Percival, W. J. et al. [The 2dFGRS Team Collabora- tion], Mon. Not. Roy. Astron. Soc. 337,1068
(2002).
\bibitem[\protect\citeauthoryear{S. Perlmutter el. al.}{1998}]{Per1} Perlmutter, S., et. al., Nature 391 (1998) 51.
\bibitem[\protect\citeauthoryear{M. Pietroni}{2003}]{piet3} Pietroni, M., Phys. Rev. D 67, 103523 (2003).

\bibitem[\protect\citeauthoryear{P. J. E. Peebles}{2003}]{pee}  Peebles, P. J. E. and  Ratra, B., Rev. Mod. Phys. 75, 559 (2003).

\bibitem[\protect\citeauthoryear{S. Perlmutter el. al.}{1999}]{per99}  Perlmutter, S.,  Turner,  M. S. and  White, M., Phys. Rev. Lett. 83, 670
(1999).

\bibitem[\protect\citeauthoryear{P. J. E. Peebles el. al.}{1988}]{peb88} Peebles, P. J. E.  and  Ratra, R., Astrophys. J.  325, L17 (1988).
\bibitem[\protect\citeauthoryear{H.V. Peiris el. al.}{2003}]{peri}  Peiris, H. V., et al., Astrophys. J. Suppl. Ser. 148, 213 (2003).

\bibitem[\protect\citeauthoryear{S. Perlmutter et al.}{1999}]{permul}  Perlmutter, S., et al., Astrophys. J. 517, 565 (1999).
\bibitem[\protect\citeauthoryear{N. Pires el. al.}{2006}]{Pires} Pires, N., Zong-Hong Zhu and J. S. Alcaniz, Phys. Rev. D 73, 123530 (2006).


\bibitem[\protect\citeauthoryear{S. Rahvar el. al.}{2007}]{sa2}  Rahvar, S.  and  Movahed, M. S.,  Phys. Rev. D  75 , 023512
(2007).

\bibitem[\protect\citeauthoryear{Raga et al.}{2002}]{raga2002} Raga, A.~C., de Gouveia Dal Pino, E.~M., Noriega-Crespo, A., Mininni, P.~D., \&
Vel{\'a}zquez, P.~F.\ 2002, A\&A, 392, 267.

\bibitem[\protect\citeauthoryear{Raga et al.}{2007}]{raga2007} Raga, A.~C., De Colle, F., Kajdi\v{c}, P., Esquivel, A., Cant\'o, J. 2007, A\&A,
465, 879.


\bibitem[\protect\citeauthoryear{B. Ratra et. al.}{1988}]{Ratra}  Ratra, B.  and  Peebles, P. J. E.,  Phys. Rev. D 37, 3406 (1988).
(2002).

\bibitem[\protect\citeauthoryear{L. Randall el. al.}{1999}]{RSII}  Randall, L.,   Sundrum, R., Phys. Rev. Lett.  83, 4690 (1999).
\bibitem[\protect\citeauthoryear{R. R. R. Reis et. al.}{2005}]{reis05}Reis R. R. R., Makler M. and Waga I., Class.Quant.Grav. 22 (2005) 353; Erratum-ibid. 22 (2005) 1191.

\bibitem[\protect\citeauthoryear{H. B. Richer el. al.}{2002}]{richer02}  Richer, H. B., et al., Astrophys. J. 574, L151 (2002).
\bibitem[\protect\citeauthoryear{A. Refregier}{2003}] {refre}  Refregier, A.,  Ann. Rev. Astron. Astrophys. 41 (2003) 645.

\bibitem[\protect\citeauthoryear{A. G. Riess el. al.}{2004}]{R04}  Riess, A. G., et al., Astrophys. J.  607, 665 (2004).

\bibitem[\protect\citeauthoryear{A.G. Riess el. al.}{1998}]{ris}  Riess, A. G., et al., Astron. J.  116, 1009 (1998).
\bibitem[\protect\citeauthoryear{Rodr{\'{\i}}guez-Mart{\'{\i}}nez et al.}{2006}]{rodriguez06} Rodr{\'{\i}}guez-Mart{\'{\i}}nez, M.,
Vel{\'a}zquez, P.~F., Binette, L., \& Raga, A.~C.\ 2006, A\& A,
448, 15.
\bibitem[\protect\citeauthoryear{Riera et al.}{2003}]{riera03} Riera, A., Garc{\'{\i}}a-Lario, P., Manchado, A., Bobrowsky, M., \& Estalella,
R.\ 2003, A\& A, 401, 1039.
\bibitem[\protect\citeauthoryear{Rybicki \& Lightman}{1979}]{rl79} Rybicki, G.~B., \& Lightman, A.~P.\ 1979, in Radiative Processes in Astrophysics, New
York, Wiley-Interscience, pp.~159-163.

\bibitem[\protect\citeauthoryear{D. H. Rudd et. al.}{2007}]{douglas07} Rudd  Douglas H., Zentner  Andrew R. and Kravtsov  Andrey V., arXiv:astro-ph/0703741.

\bibitem[\protect\citeauthoryear{V. Sahni el. al.}{2000}]{sa00} Sahni,V. and Starobinsky, A. Int. J. Mod. Phys. D  9, 373 (2000).

\bibitem[\protect\citeauthoryear{V. Sahni el. al.}{2002}]{Sah1}  Sahni, V. and  Shtanov, Y., Int. J. Mod. Phys. D 11, 1515 (2002)
.

\bibitem[\protect\citeauthoryear{V. Sahni et. al.}{2003}]{Sah2} Sahni, V.  and Shtanov, Y., J. Cosmol. Astropart. Phys. 11 (2003)
014.
\bibitem[\protect\citeauthoryear{U. Seljak et. al.}{2004}] {Selj} Seljak, U., et. al., Phys.Rev. D71 (2005) 103515.
\bibitem[\protect\citeauthoryear{U. Seljak and M. Zaldarriaga}{1999}]{seljak98}  Seljak, U. and  Zaldarriaga, M., Phys.Rev.Lett. 82 (1999) 2636-2639.




\bibitem[\protect\citeauthoryear{A. Sheykhi el. al.}{2007a}]{She1} Sheykhi, A.,   Wang, B. and
 Cai, R. G.,  Nucl. Phys. B, 779 (2007a) 1.
\bibitem[\protect\citeauthoryear{A. Sheykhi el. al.}{2007b}]{Sheykhi} Sheykhi, A.,  B. Wang and N. Riazi, Phys. Rev. D  75, 123513 (2007b).

\bibitem[\protect\citeauthoryear{A. Sheykhi, B. Wang and R.G. Cai}{2007c}]{She20}Sheykhi, A., Wang, B.  and
Cai, R. G.,  Phys. Rev. D 76 (2007c) 023515.
\bibitem[\protect\citeauthoryear{D. N. Spergel el. al.}{2003a}]{spe03}  Spergel, D. N.,   Verde, L.,  Peiris, H. V., et al., The Astrophysical Journal Supplement Series, (2003), Volume 148, Issue 1, pp. 175-194
.
\bibitem[\protect\citeauthoryear{H. Spinrard }{1997}]{spin}  Spinrard, H.,  Astrophys. J.  484, 581
(1997).
\bibitem[\protect\citeauthoryear{B. P. Schmidt el. al.}{1998}]{Schmidt}  Schmidt, B. P., et al., Astrophys. J.  507, 46 (1998).

\bibitem[\protect\citeauthoryear{Soker \& Bisker}{2006}]{soker06} Soker, N., \& Bisker, G.\ 2006, MNRAS, 369,
1115.
\bibitem[\protect\citeauthoryear{P. J. Steinhardt}{1999}]{stei}  Steinhardt, P. J.,  Wang, L.  and
Zlatev, I.,  Phys. Rev. D 59, 123504 (1999).


\bibitem[\protect\citeauthoryear{M. Tegmark, et al.}{2004a}]{tegmark104} Tegmark, M. et al. (the SDSS collaboration), Phys. Rev. D
{\bf 69}, 103501 (2004a).

\bibitem[\protect\citeauthoryear{M. Tegmark, R. Michael Blanton et al.}{2004b}]{tegmark204} Tegmark, M. et al. (the SDSS collaboration), Astrophys. J. {\bf 606}, 702
(2004b).

\bibitem[\protect\citeauthoryear{M. Tegmark, et al.}{2002}]{tegmark02}  Tegmark, M., Hamilton, A. J. S. and  Xu, Y., Mon.
Not. R. Astron. Soc. {\bf 335}, 887 (2002).




\bibitem[\protect\citeauthoryear{The Gold dataset}{2006}]{new} The Gold dataset is available at http://braeburn.pha.jhu.edu/˜ariess/R06.
\bibitem[\protect\citeauthoryear{J. L. Tonry el. al.}{2003}]{Tonry}  Tonry, J. L., et al., Astrophys. J.  594, 1 (2003).
\bibitem[\protect\citeauthoryear{ M. S. Turner et. al.}{1997}]{Turn}  Turner, M. S. and  White, M.,  Phys. Rev. D 56, R
4439 (1997).
\bibitem[\protect\citeauthoryear{ D. F. Torres}{2002}]{Torres} Torres, D. F., Phys. Rev. D
66, 043522 (2002).
\bibitem[\protect\citeauthoryear{Tafoya et al.}{2007}]{tafoya07} Tafoya, D., et al.\ 2007, AJ, 133,
364.

\bibitem[\protect\citeauthoryear{K. I. Umezu et. al.}{2006}]{Umezu}  Umezu, K. I.,  Ichiki,   K.,  Kajino, T.,  Mathews,  G. J.,
 Nakamura, R., and  Yahiro, M., Phys.
Rev. D 73  (2006) 063527.
\bibitem[\protect\citeauthoryear{Van Leer}{1982}]{vanleer1982} Van Leer, B., ICASE Report No. 82-30 (1982).
\bibitem[\protect\citeauthoryear{B. Wang et. al.}{2006a}]{Wang1}  Wang, B.,  Gong, Y. and   Abdalla, E., Phys. Lett.  B 624, (2006a)
141.

\bibitem[\protect\citeauthoryear{B. Wang et. al}{2006b}]{Wang2}  Wang, B.,  Lin, C. Y. and   Abdalla, E.,  Phys. Lett. B 637, (2006b) 357.

\bibitem[\protect\citeauthoryear{B. Wang el. al.}{2006}]{Wang3}  Wang, B.,  Zang, J. D.,  Lin, C.Y.,  Abdalla, E. and S. Micheletti, astro-ph/0607126.
\bibitem[\protect\citeauthoryear{S. Weinberg}{1989}]{wein}  Weinberg, S.,  Rev. Mod. Phys. 61, 1 (1989).
\bibitem[\protect\citeauthoryear{D. H. Weinberg el. al.}{2005}]{Wein2}  Weinberg, D. H.,  New Astron. Rev. 49 (2005) 337.
\bibitem[\protect\citeauthoryear{C. Wetterich}{1998}]{wet88}  Wetterich, C.,  Nucl. Phys. B 302, 668 (1988).

\bibitem[\protect\citeauthoryear{L. Wang el. al.}{2000}]{wan00}  Wang, L.,  Caldwell,  R. R.,  Ostriker,  J. P. and
Steinhardt, P. J., Astrophys. J.  530, 17 (2000).
\bibitem[\protect\citeauthoryear{Zhang}{1995}]{zhang95} Zhang, C.~Y.\ 1995, ApJS, 98,
659.
\bibitem[\protect\citeauthoryear{ X. Zhang}{2005}]{zhang5} Zhang, X., Phys. Lett. B 611, 1 (2005).
\bibitem[\protect\citeauthoryear{ I. Zlatev}{1999}]{Zlate} Zlatev, I., Wang, L. and
Steinhardt, P. J., Phys. Rev. Lett. 82, 896 (1999).

\bibitem[\protect\citeauthoryear{X. Zhang el. al.}{2005}]{zang} Zhang, X.  and F.Q. Wu, Phys. Rev. D  72, 043524
(2005).















\end{thebibliography}
\end{document}